\documentclass[10pt,a4paper,twoside]{article}
\usepackage{epsfig}
\usepackage{baltlat6}
\usepackage{array}
\usepackage{dcolumn}
\usepackage{longtable}
\usepackage{lscape}
\begin{document}
\ \
\vspace{0.5mm}
\setcounter{page}{1}
\vspace{8mm}

\titlehead{Baltic Astronomy, vol.\,19, 1--33, 2010}

\titleb{YOUNG STARS IN THE CAMELOPARDALIS DUST AND\\ MOLECULAR CLOUDS.
VI. YSOs VERIFIED BY SPITZER AND\\ AKARI INFRARED PHOTOMETRY}

\begin{authorl}
\authorb{V. Strai\v{z}ys}{} and
\authorb{A. Kazlauskas}{}
\end{authorl}

\moveright-3.2mm
\vbox{
\begin{addressl}
\addressb{}{Institute of Theoretical Physics and Astronomy, Vilnius
University,\\
  Go\v{s}tauto 12, Vilnius LT-01108, Lithuania}
\end{addressl}
}

\submitb{Received 2010 April 1; accepted 2010 April 25}

\begin{summary} Using photometric data of infrared surveys, young
stellar object (YSO) status is verified for 141 objects selected in our
previous papers in the Cassiopeia and Camelopardalis segment of the
Milky Way bounded by Galactic coordinates ($\ell$, $b$) =
(132--158\degr, $\pm$12\degr).  The area includes the known star-forming
regions in the emission nebulae W3, W4 and W5 and the massive YSO
AFGL\,490.  Spectral energy distribution (SED) curves between 700 nm and
160 $\mu$m, constructed from the GSC~2, 2MASS, IRAS, MSX, Spitzer and
AKARI data, are used to estimate the evolutionary stages of these stars.
We confirm the YSO status for most of the objects.  If all of the
investigated objects were YSOs, 45\% of them should belong to Class I,
41\% to class II and 14\% to Class III.  However, SEDs of some of these
objects can be affected by nearby extended infrared sources, like
compact H\,II regions, infrared clusters or dusty galaxies.
\end{summary}


\resthead{Young stars in the Camelopardalis dust and molecular clouds.
VI.}{V. Strai\v{z}ys, A. Kazlauskas}

\sectionb{1}{INTRODUCTION}

In the previous papers in this series, II and III (Strai\v{z}ys \&
Laugalys 2007b, 2008a), the Camelopardalis segment of the Milky Way,
bounded by Galactic coordinates ($\ell$,\,$b$) = (132--158\degr,
$\pm$\,12\degr), was shown to be an active region of star formation.
Apart from Camelopardalis, the area includes also the neighboring
regions of Cassiopeia (with H\,II regions W3, W4 and W5), Perseus and
Auriga.  Applying infrared photometry from the 2MASS, IRAS and MSX
databases, we identified about 190 objects exhibiting photometric
properties of young stellar objects (YSOs).  In the present paper, these
objects will be called as SL objects.  In addition, the area contains
more than 40 massive stars of the Cam OB1 association and about 20 young
lower-mass stars which exhibit emission in H$\alpha$ or belong to a
class of irregular variable stars of types IN and IS (Strai\v{z}ys \&
Laugalys 2007a, Paper I).  The region contains a high-mass pre-stellar
object, AFGL\,490, embedded in the densest part of the dust cloud
TGU\,942 (Dobashi et al. 2005).  According to Zdanavi\v{c}ius et al.
(2005) and Strai\v{z}ys \& Laugalys (2007a, 2008b), star forming is most
active in two layers of dust/molecular clouds at the distances 150--300
pc and 800--900 pc, which belong to the Gould Belt and the Cam~OB1
association, respectively.  A large fraction of these objects belong
also to the Perseus arm which in this direction is located at a distance
of 2.0--2.5 kpc.

In an attempt to verify the YSO status of SL objects, we obtained
far-red spectra for 30 brightest objects (Corbally et al. 2008, 2009;
Papers IV and V).  All of them exhibited emission in H$\alpha$ line and
some also in Ca\,II and O\,I lines, in agreement with their
pre-main-sequence status.  This was also confirmed by infrared spectral
energy distributions (SEDs) where observations by IRAS and MSX missions
were available.

In 2007--2010, a few papers describing Spitzer observations in the IRAC
passbands and 24~$\mu$m MIPS passband in star-forming regions (SFRs) of
this area were published.  Recently, the first two catalogs based on
observations in the middle and far infrared from the Japanese AKARI
satellite (Murakami et al. 2007) have become
available.{\footnote{~http://www.ir.isas.jaxa.jp/AKARI/Observation/PSC/Public/}
In the present paper we describe our search of the suspected YSOs in the
Spitzer and AKARI data, the construction of their SEDs using the GSC~2,
2MASS, IRAS, MSX, Spitzer and AKARI data, and the estimation of
evolutionary stages of the confirmed YSOs.

\sectionb{2}{IDENTIFICATION OF SL OBJECTS IN THE AKARI, SPITZER AND\\
IRAS CATALOGS}
\vskip2mm

The AKARI data released for public use contain two all-sky catalogs:
the IRC Point Source Catalog and the FIS Bright Source Catalog.  The
first one contains the results of photometry of about 871\,000 point
sources in two mid-infrared passbands with the mean wavelengths 9~$\mu$m
and 18~$\mu$m (Kataza et al. 2010; Ishihara et al. 2010).  The second
catalog contains the results of photometry of about 427\,000 sources in
the far-infrared passbands with the mean wavelengths 65, 90, 140 and 160
$\mu$m (Shirahata et al. 2009; Yamamura et al. 2010).

In the AKARI catalogs, we identified 104 SL objects by comparing their
equatorial coordinates given in the 2MASS and AKARI catalogs.  The
search radius around the positions of SL objects (taken from 2MASS) was
10\arcsec\ for the IRC catalog and 1\arcmin\ for the FIS catalog.  In
most cases, only one AKARI object was present within the above-mentioned
radii around the SL objects. 42 objects were found both in the IRC and
FIS catalogs, 28 objects only in IRC and 32 objects only in FIS.  Fluxes
are available not in all passbands of the IRC and FIS surveys; only
reliable fluxes with the quality flag 3 were taken.

The IRAS objects were identified through a search in a 2\arcmin\ radius
circle around the SL/2MASS objects using the Vizier facility of the CDS.
In some cases the IRAS and 2MASS objects can be different.  Such cases
will be discussed in Section 4.

Table 1 gives the identifications of SL objects in the 2MASS, IRAS and
AKARI IRC and FIS catalogs.  The angular distances of the IRAS and AKARI
objects from the 2MASS positions are also given.

The Spitzer data for SL objects were taken from the following sources:
Gutermuth et al.  (2009) for 11 objects in the AFGL\,490 SFR,
Ruch et al.  (2007) for 5 objects in the W3 SFR, Koenig et al.
(2008) for 21 objects in the W5 SFR and Kumar Dewangan \& Anandarao
(2010) for one object in the AFGL\,437 SFR.

 \noindent
\landscape
\extrarowheight=-.9pt
\tabcolsep=7pt
\small
\centerline{\parbox{160mm}{\baselineskip=9pt
{\smallbf\ \ Table 1.}{\small\
Identification of the SL objects in the 2MASS, IRAS and AKARI
catalogs having mid- and far-infrared measurements.  The objects without
IRAS and AKARI numbers have been measured only by Spitzer and MSX. Nine
known YSOs in the same area are listed at the end of the Table.
Separations between the 2MASS and the IRAS and AKARI objects are given
in arcseconds.}}}
\vskip-2mm
\begin{longtable}{lccrcrcr}
\hline
 \huad ID      &       2MASS \hstrut &    IRAS     &  Sep. &   AKARI-IRC-V1   &  Sep. &   AKARI-FIS-V1   &  Sep. \\
\noalign{\vspace{1mm}}
\hline
\noalign{\vspace{1mm}}
\endfirsthead
\multicolumn{8}{l}{{\smallbf\ \ Table 1.}{\small\ Continued\lstrut}}\\
\hline
\huad ID        &       2MASS\hstrut &    IRAS     &  Sep. &   AKARI-IRC-V1   &  Sep. &   AKARI-FIS-V1   &  Sep. \\
\noalign{\vspace{1mm}}
\hline
\noalign{\vspace{1mm}}
\endhead
\endfoot
SL\,1           &  J02115015+6239356 &  02081+6225 &   5.3 &  J0211497+623941 &   6.8 &  J0211493+623952 &  17.6 \\
SL\,3           &  J02413774+6947098 &  02371+6934 &   1.2 &  J0241377+694708 &   1.0 &  J0241387+694707 &   5.7 \\
SL\,4           &  J02444712+6959571 &  02402+6947 &   9.3 &  J0244472+695957 &   0.8 &                  &       \\
SL\,5           &  J02474082+6933066 &  02432+6919 &  91.5 &                  &       &                  &       \\
SL\,6           &  J02465383+6906314 &             &       &  J0246539+690632 &   0.7 &                  &       \\
SL\,7           &  J02165920+6108158 &  02134+6053 & 103.2 &                  &       &  J0217020+610836 &  28.0 \\
SL\,8           &  J02513724+6914018 &  02470+6901 &  14.2 &                  &       &                  &       \\
SL\,9           &  J02202209+6132219 &  02167+6118 &  24.8 &  J0220220+613221 &   0.6 &  J0220240+613155 &  29.8 \\
SL\,10           &  J02192184+6107069 &  02157+6053 &  21.7 &  J0219216+610707 &   1.1 &  J0219252+610721 &  28.3 \\
SL\,11           &  J02210026+6128560 &  02173+6113 & 116.2 &                  &       &                  &       \\
SL\,13           &  J02230386+6205135 &             &       &                  &      &  J0223013+620445  &  34.0 \\
SL\,14           &  J02541453+6918367 &             &       &                  &       &                  &       \\
SL\,15           &  J02230616+6158437 &             &       &  J0223061+615843 &   0.2 &                  &       \\
SL\,16           &  J02210449+6106045 &  02174+6052 &  29.7 &  J0221044+610604 &   0.6 &  J0221063+610602 &  13.1 \\
SL\,18           &  J02245191+6209367 &             &       &                  &       &  J0224555+620941 &  25.6 \\
SL\,20           &  J02250325+6159477 &             &       &                  &       &                  &       \\
SL\,21           &  J02231856+6125416 &             &       &  J0223185+612541 &   0.3 &                  &       \\
SL\,23           &  J02253880+6207297 &             &       &                  &       &  J0225340+620701 &  44.2 \\
SL\,25           &  J02255757+6211497 &             &       &                  &       &  J0225598+621159 &  18.2 \\
SL\,27           &  J02254436+6206117 &  02219+6152 &  13.0 &  J0225442+620616 &   5.3 &  J0225467+620626 &  22.1 \\
SL\,30           &  J02253198+6157249 &             &       &                  &       &                  &       \\
SL\,31           &  J02262447+6210384 &             &       &                  &       &                  &       \\
SL\,34           &  J02262572+6203559 &  02226+6150 &   5.7 &  J0226254+620353 &   2.8 &                  &       \\
SL\,35           &  J02265368+6212000 &  02230+6157 &  81.0 &                  &       &  J0226509+621218 &  26.5 \\
SL\,36           &  J02263669+6153085 &             &       &                  &       &  J0226313+615306 &  38.0 \\
SL\,37           &  J02271602+6200506 &             &       &  J0227172+620053 &   8.8 &  J0227184+620030 &  26.8 \\
SL\,38           &  J02254737+6120562 &  02220+6107 &   9.6 &  J0225475+612056 &   1.4 &  J0225481+612057 &   5.4 \\
SL\,39           &  J02264501+6129430 &             &       &  J0226452+612942 &   1.5 &  J0226467+612933 &  15.8 \\
SL\,40           &  J02271217+6137055 &             &       &  J0227115+613702 &   5.7 &  J0227112+613726 &  21.1 \\
SL\,41           &  J02252125+6057581 &  02215+6044 &  21.1 &  J0225210+605758 &   1.8 &  J0225200+605809 &  14.0 \\
SL\,43           &  J02282010+6130351 &  02244+6117 &  56.8 &                  &       &                  &       \\
SL\,44           &  J02282148+6128369 &  02245+6115 &   7.9 &                  &       &  J0228217+612856 &  19.3 \\
SL\,48           &  J02363576+6031518 &  02327+6019 &  24.8 &  J0236354+603153 &   2.9 &  J0236341+603214 &  25.7 \\
SL\,49           &  J02512318+6209031 &  02475+6156 &  75.4 &  J0251228+620903 &   2.5 &  J0251303+620847 &  52.3 \\
SL\,50           &  J02470551+6102451 &  02431+6050 &   5.9 &  J0247053+610245 &   1.5 &  J0247059+610249 &   5.0 \\
SL\,51           &  J02460197+6039438 &             &       &                  &       &  J0246010+604031 &  48.2 \\
SL\,52           &  J02482571+6055051 &  02445+6042 &   4.5 &  J0248255+605506 &   1.8 &  J0248248+605508 &   6.8 \\
SL\,53           &  J02485350+6045115 &             &       &                  &       &                  &       \\
SL\,54           &  J02492991+6047287 &  02455+6034 &  56.2 &  J0249297+604728 &   0.9 &                  &       \\
SL\,55           &  J02492190+6045041 &             &       &                  &       &                  &       \\
SL\,56           &  J02494739+6042090 &  02459+6029 &   2.6 &  J0249472+604208 &   1.1 &  J0249447+604238 &  34.7 \\
SL\,57           &  J02502403+6036287 &  02465+6024 &  27.8 &                  &       &                  &       \\
SL\,58           &  J02590280+6225295 &             &       &  J0259026+622530 &   1.6 &                  &       \\
SL\,59           &  J02512557+6006048 &             &       &                  &       &                  &       \\
SL\,60           &  J02513283+6003542 &  02476+5950 &  49.5 &  J0251328+600353 &   0.7 &  J0251330+600341 &  13.0 \\
SL\,61           &  J02551296+6042404 &             &       &  J0255129+604241 &   0.6 &  J0255116+604258 &  20.7 \\
SL\,62           &  J02572154+6053097 &  02532+6042 & 106.9 &  J0257214+605309 &   1.1 &  J0257225+605314 &   8.2 \\
SL\,63           &  J02572160+6041198 &  02534+6029 &  28.0 &                  &       &  J0257171+604203 &  54.2 \\
SL\,64           &  J02573189+6038171 &  02533+6026 & 103.5 &                  &       &  J0257323+603829 &  12.6 \\
SL\,65           &  J03001592+6059240 &  02563+6047 &  33.2 &  J0300158+605924 &   0.6 &  J0300155+605934 &  10.6 \\
SL\,66           &  J03013187+6029256 &  02575+6017 &  14.3 &                  &       &  J0301314+602926 &   3.9 \\
SL\,68           &  J03113336+6222086 &  03074+6211 &  51.5 &                  &       &                  &       \\
SL\,69           &  J03025804+6027571 &             &       &                  &       &                  &       \\
SL\,70           &  J03031387+6028092 &             &       &                  &       &                  &       \\
SL\,71           &  J03032084+6029284 &             &       &  J0303209+602928 &   0.7 &                  &       \\
SL\,72           &  J03031615+6027502 &  02593+6016 &  12.7 &                  &       &                  &       \\
SL\,74           &  J03051264+6048430 &             &       &  J0305124+604844 &   2.0 &  J0305110+604915 &  34.0 \\
SL\,75           &  J03032586+6023095 &  02595+6011 &  27.7 &  J0303257+602309 &   1.3 &                  &       \\
SL\,77           &  J02512410+5542038 &  02476+5529 &  72.2 &                  &       &  J0251208+554217 &  30.7 \\
SL\,78           &  J02514696+5542014 &             &       &  J0251469+554202 &   1.3 &  J0251409+554223 &  55.8 \\
SL\,79           &  J03104626+5930035 &  03068+5918 &  21.7 &  J0310461+593003 &   1.0 &                  &       \\
SL\,80           &  J03072452+5830433 &  03035+5819 &  12.2 &  J0307241+583047 &   5.1 &  J0307233+583112 &  30.2 \\
SL\,81           &  J03153845+6002404 &  03116+5951 &  14.0 &  J0315382+600241 &   1.4 &                  &       \\
SL\,82           &  J03172590+6009417 &  03134+5958 &   5.6 &  J0317258+600942 &   0.8 &  J0317256+600944 &   3.1 \\
SL\,83           &  J03152500+5855482 &  03114+5844 &   9.9 &  J0315249+585548 &   0.4 &  J0315254+585557 &   9.4 \\
SL\,84           &  J03153715+5857208 &             &       &                  &       &  J0315306+585712 &  51.2 \\
SL\,85           &  J03300237+6125473 &             &       &  J0330022+612547 &   0.8 &                  &       \\
SL\,86           &  J03130254+5804483 &  03091+5753 &  11.1 &  J0313026+580449 &   1.3 &  J0313029+580443 &   5.8 \\
SL\,87           &  J03084929+5649572 &  03050+5638 &  12.8 &  J0308489+564957 &   2.9 &  J0308480+565001 &  11.2 \\
SL\,88           &  J03081580+5623388 &  03045+5612 &  15.7 &                  &       &  J0308185+562340 &  22.7 \\
SL\,89           &  J03264934+5845238 &  03228+5834 &   6.4 &  J0326493+584525 &   1.2 &  J0326475+584524 &  14.5 \\
SL\,91           &  J03265964+5842199 &             &       &                  &       &                  &       \\
SL\,93           &  J03271645+5844376 &  03233+5833 &  79.0 &  J0327175+584440 &   9.2 &                  &       \\
SL\,94           &  J03273489+5847485 &             &       &                  &       &                  &       \\
SL\,95           &  J03273876+5847000 &  03236+5836 &   2.6 &  J0327386+584700 &   0.7 &                  &       \\
SL\,96           &  J03271170+5840269 &             &       &  J0327114+584028 &   2.4 &  J0327157+583959 &  41.8 \\
SL\,97           &  J03280189+5847091 &             &       &                  &       &                  &       \\
SL\,98           &  J03274898+5812164 &             &       &  J0327489+581218 &   1.7 &                  &       \\
SL\,100          &  J03310105+5803448 &             &       &                  &       &  J0330594+580354 &  16.3 \\
SL\,101          &  J03290756+5701336 &             &       &  J0329074+570133 &   0.6 &                  &       \\
SL\,102          &  J03330101+5734127 &  03290+5724 &  13.4 &  J0333009+573416 &   3.4 &  J0333018+573412 &   6.4 \\
SL\,104          &  J03224839+5503371 &  03188+5452 &  82.9 &                  &       &                  &       \\
SL\,106          &  J03235137+5452434 &  03201+5442 &  33.2 &  J0323516+545240 &   3.7 &                  &       \\
SL\,107          &  J03341497+5653213 &  03303+5643 &  18.1 &                  &       &                  &       \\
SL\,108          &  J03250185+5456160 &  03211+5446 &  72.5 &                  &       &                  &       \\
SL\,109          &  J03331001+5510549 &  03293+5500 &   3.6 &  J0333098+551054 &   1.1 &  J0333100+551054 &   1.4 \\
SL\,110          &  J03323036+5450449 &  03286+5440 &   8.5 &                  &       &  J0332296+545051 &   8.8 \\
SL\,111          &  J03364225+5447431 &  03328+5437 &  11.2 &  J0336423+544743 &   0.8 &                  &       \\
SL\,112          &  J03390466+5345239 &  03353+5333 & 148.1 &  J0339047+534524 &   0.8 &                  &       \\
SL\,113          &  J03484016+5432136 &  03447+5422 &  27.5 &                  &       &  J0348418+543158 &  21.4 \\
SL\,115          &  J03554277+5350074 &  03518+5341 &  26.6 &  J0355425+535007 &   2.2 &                  &       \\
SL\,116          &  J03553004+5345419 &  03516+5336 &  17.3 &  J0355299+534542 &   0.8 &  J0355278+534520 &  29.5 \\
SL\,118          &  J04082550+5508284 &  04044+5500 &  19.1 &  J0408254+550829 &   1.0 &  J0408260+550834 &   6.9 \\
SL\,120          &  J04075001+5441157 &  04038+5433 &  39.2 &  J0407497+544116 &   2.5 &  J0407499+544129 &  13.1 \\
SL\,121          &  J04192876+5418579 &             &       &  J0419287+541857 &   0.8 &                  &       \\
SL\,123          &  J04034287+5123423 &             &       &  J0403428+512343 &   0.8 &                  &       \\
SL\,124          &  J04044961+5126572 &  04010+5118 &  23.4 &  J0404496+512657 &   0.2 &  J0404500+512658 &   3.7 \\
SL\,125          &  J04145692+5220313 &  04110+5212 &  11.9 &                  &       &  J0414574+522036 &   6.2 \\
SL\,126          &  J04193181+5251221 &  04156+5244 &  10.3 &  J0419326+525121 &   7.7 &  J0419329+525122 &  10.0 \\
SL\,129          &  J04111220+5110238 &  04073+5102 &  63.1 &                  &       &                  &       \\
SL\,130          &  J04101185+5059544 &  04064+5052 &   3.5 &  J0410118+505954 &   0.3 &  J0410117+505942 &  12.2 \\
SL\,131          &  J04071005+5018240 &  04034+5010 &   4.5 &  J0407100+501823 &   0.1 &  J0407097+501824 &   3.5 \\
SL\,132          &  J04083557+5031588 &             &       &  J0408356+503158 &   0.3 &                  &       \\
SL\,133          &  J04154142+4915218 &  04119+4907 &   5.3 &                  &       &  J0415418+491509 &  13.4 \\
SL\,134          &  J04232978+4919314 &  04198+4912 &  36.8 &                  &       &  J0423303+491923 &  10.3 \\
SL\,135          &  J04344416+5039275 &  04308+5033 &  23.7 &                  &       &  J0434442+503945 &  17.3 \\
SL\,137          &  J04271337+4804149 &  04235+4757 &   2.9 &  J0427135+480415 &   1.8 &  J0427128+480419 &   6.7 \\
SL\,138          &  J04264419+4642295 &  04230+4635 &  30.2 &                  &       &  J0426444+464200 &  29.2 \\
SL\,140          &  J04254952+4604387 &  04222+4557 &  27.0 &                  &       &  J0425502+460431 &  10.4 \\
SL\,141          &  J04204895+4417280 &  04172+4411 &  82.2 &                  &       &                  &       \\
SL\,142          &  J04022981+4042418 &  03591+4034 &  10.3 &  J0402298+404241 &   0.2 &                  &       \\
SL\,143          &  J03262717+5848241 &             &       &                  &       &  J0326263+584818 &   8.9 \\
SL\,145          &  J03270030+5846164 &             &       &                  &       &                  &       \\
SL\,146          &  J03270077+5844307 &             &       &                  &       &                  &       \\
SL\,147          &  J03265892+5842531 &             &       &                  &       &                  &       \\
SL\,148          &  J03272257+5843170 &  03233+5833 &  23.4 &                  &       &  J0327253+584336 &  28.1 \\
SL\,149          &  J03280190+5847030 &             &       &                  &       &                  &       \\
SL\,155          &  J03273738+5803381 &  03238+5752 &  79.1 &                  &       &                  &       \\
SL\,158          &  J03300545+5813253 &  03261+5803 &   1.0 &  J0330054+581325 &   0.4 &  J0330052+581327 &   2.9 \\
SL\,159          &  J03295426+5805396 &  03260+5755 &  60.3 &                  &       &                  &       \\
SL\,160          &  J03285987+5753402 &             &       &                  &       &  J0329039+575302 &  49.6 \\
SL\,162          &  J03312962+5816425 &             &       &                  &       &  J0331284+581716 &  35.3 \\
SL\,163          &  J03325315+5827511 &  03289+5818 &  91.0 &                  &       &                  &       \\
SL\,165          &  J03340073+5816376 &             &       &  J0334004+581637 &   2.0 &                  &       \\
SL\,169          &  J03333964+5809204 &  03298+5758 &  84.7 &                  &       &                  &       \\
SL\,171          &  J03313040+5721567 &             &       &  J0331302+572157 &   1.6 &                  &       \\
SL\,172          &  J03334575+5732396 &  03296+5723 &  85.2 &                  &       &                  &       \\
SL\,174          &  J03300550+5547529 &  03262+5536 &  80.9 &                  &       &                  &       \\
SL\,175          &  J03205334+5849395 &  03167+5840 & 101.4 &  J0320530+584940 &   2.4 &  J0320529+584947 &   8.3 \\
SL\,176          &  J03172447+5754136 &  03135+5743 &   5.3 &  J0317244+575412 &   1.1 &                  &       \\
SL\,177          &  J03275850+5858341 &  03239+5849 &  60.4 &                  &       &                  &       \\
SL\,178          &  J03235621+5804391 &  03199+5755 &  90.0 &                  &       &                  &       \\
SL\,179          &  J03233167+5757520 &  03194+5746 &  84.7 &                  &       &                  &       \\
SL\,180          &  J03251924+5811455 &  03213+5801 &  22.0 &  J0325191+581146 &   1.2 &                  &       \\
SL\,181          &  J03300680+5826389 &  03263+5816 &  91.6 &                  &       &  J0330097+582727 &  53.3 \\
SL\,182          &  J03265911+5740027 &  03231+5730 &  78.5 &                  &       &  J0326583+573959 &   7.0 \\
SL\,183          &  J03284786+5755560 &  03248+5745 &   2.6 &  J0328479+575557 &   1.0 &  J0328478+575552 &   4.0 \\
SL\,184          &  J03300294+5805348 &  03260+5755 &   9.2 &  J0330027+580535 &   1.6 &                  &       \\
SL\,185          &  J03312047+5805488 &  03275+5755 &  94.2 &                  &       &  J0331217+580626 &  38.8 \\
SL\,186          &  J03330645+5817165 &  03292+5806 &  47.3 &                  &       &                  &       \\
SL\,187          &  J03354115+5725589 &  03317+5716 &  53.3 &                  &       &                  &       \\
\noalign{\vspace{1mm}}
\end{longtable}
\nopagebreak
\vskip-1mm

\begin{tabular}{lccrcrcr}
\multicolumn{8}{l}{{\smallbf\ \ Table 1.}{\small\ Continued\lstrut}}\\
\hline
\hhuad\huad ID   &       2MASS\hstrut &    IRAS     &  Sep. &   AKARI-IRC-V1   &  Sep. &   AKARI-FIS-V1   &  Sep. \\
\noalign{\vspace{1mm}}
\hline
\noalign{\vspace{1mm}}
KW97\,14-24       & J03012159+6028566 & 02575+6017 &  80.5 &                  &       &                  &     \\
KW97\,14-52       & J03082443+5943265 &            &       &  J0308244+594326 &   0.4 &                  &     \\
IRAS\,03243+5819  & J03281460+5829374 & 03243+5819 &  18.4 &  J0328144+582938 &   1.3 &                  &     \\
LkH$\alpha$\,272  & J03490512+3856172 & 03456+3846 &  72.9 &  J0349050+385617 &   0.8 &                  &     \\
XY\,Per           & J03493638+3858556 & 03462+3849 &   2.8 &  J0349363+385855 &   0.1 &  J0349363+385902 & 6.0 \\
Gahm\,21          & J03595342+5131463 & 03561+5123 &  30.9 &                  &       &                  &     \\
KW97\,16-55       & J04011993+5311221 & 03573+5302 & 100.2 &                  &       &                  &     \\
OS\,Per           & J04362931+4916207 & 04325+4911 & 113.0 &                  &       &                  &     \\
V347\,Aur         & J04565702+5130509 & 04530+5126 &   2.2 &  J0456570+513050 &   0.4 &  J0456569+513050 & 1.1 \\
\noalign{\vspace{1mm}}
\hline
\end{tabular}
\endlandscape

\sectionb{3}{SPECTRAL ENERGY DISTRIBUTIONS}

For the identified SL objects a table with reliable fluxes from the
2MASS, IRAS, MSX, Spitzer and AKARI surveys was compiled.  The data in
the table were supplemented by the photographic red $F$ magnitudes
extracted from the GSC~2.3.2 catalog available at the CDS (Lasker et al.
2008).  The IRAS, MSX and AKARI fluxes in Janskys were transformed
simply to $\log \lambda F_{\lambda}$.  The $F$ and 2MASS magnitudes were
transformed to $\log \lambda F_{\lambda}$ by the equations given in
Paper II (p.\,341--342), accepting that the magnitudes $F$ = $R$ and
using for them the absolute calibration from Strai\v zys (1992).

The Vega fluxes in Janskys, used in the transformation of the Spitzer
magnitudes to fluxes, were taken from Reach et al.  (2005) for the IRAC
data and from Rieke et al.  (2008) for the 24 $\mu$m MIPS data:
\begin{equation}
[3.6] = -2.5 \log\,(F[Jy]/280.9),
\vspace{-1mm}
\end{equation}
\begin{equation}
[4.5] = -2.5 \log\,(F[Jy]/179.7),
\end{equation}
\begin{equation}
[5.8] = -2.5 \log\,(F[Jy]/115.0),
\end{equation}
\begin{equation}
[8.0] = -2.5 \log\,(F[Jy]/64.1),~
\end{equation}
\begin{equation}
[24.0] = -2.5 \log\,(F[Jy]/7.15).
\end{equation}

The resulting values of $\log \lambda F_{\lambda}$ for 141 SL objects
are given in Tables 2 and 3:  the first one contains 127 objects for
which the data from IRAS, MSX or AKARI surveys are available, and the
second one contains 37 objects identified in the Spitzer catalogs.  At
the end of Table 2 the data for nine known YSOs in the same area,
discussed in Papers I and IV, are added.  The plots of $\log \lambda
F_{\lambda}$ vs. log\,$\lambda$ for the 141 SL objects and  nine
known YSOs are shown in Figure~1.

The traditional classification scheme of YSOs (Lada 1987) is based on
their SEDs longward of 2~$\mu$m. The spectral index
\begin{equation}
a = {{d\log\,(\lambda F_{\lambda})} \over {d\log\,\lambda}}
\end{equation}
for Class I objects is positive, for Class II objects it is close to
zero or slightly negative and for Class III objects it is negative.
Evolutionary interpretation of SEDs has been given by Whitney et al.
(2003a,b, 2004) and Robitaille et al.  (2006, 2007) using tens
of thousands of YSO models with disks and envelopes.

Examination of our $\log \lambda F_{\lambda}$ vs. log\,$\lambda$ plots
indicates that SEDs longward of 2~$\mu$m for 45\% of the objects show
increase in intensity, for 41\% of the objects SEDs remain more or less
at the same level and for 14\% of the objects SEDs exhibit a tendency to
decrease.  If all of these objects were YSOs, then we
should conclude that the majority of them are young stars either
embedded into dense gas and dust envelopes which reradiate energy of the
central source (Class I) or have optically thick disks and possible
remains of tenuous infalling envelopes (Class II, T Tauri and Herbig
Ae/Be stars).  Only a small fraction of the objects with the SEDs with a
negative spectral index should belong to Stage III (or post T
Tauri-type) objects with thin disks and weak emission lines.
These objects hardly can be normal heavily reddened stars since SEDs of
such stars exhibit much steeper negative slope (see Paper II, Figure
5).

\landscape
\footnotesize
\noindent
\extrarowheight=-.97pt
\tabcolsep=3pt
\begin{longtable}{rD..{3.3}D..{3.3}D..{3.3}D..{3.3}D..{3.3}D..{3.3}D..{3.3}D..{3.3}
D..{3.3}D..{3.3}D..{3.3}D..{3.3}D..{3.3}D..{3.3}D..{3.3}}
\noalign{\parbox[c]{178mm}{\baselineskip=9pt
{\smallbf\ \ Table 2.}{\small\ The values of
$\log \lambda F_{\lambda}$ for the suspected YSOs
in the passbands of the 2MASS, IRAS, MSX and AKARI surveys. $F$ are red
photographic magnitudes from the GSC~2.3.2 catalog. \lstrut}}}
\hline
SL&\multicolumn{1}{c}{$F$}&\multicolumn{1}{c}{$J$}&\multicolumn{1}{c}{$H$}&\multicolumn{1}{c}{$K$}\hstrut&
\multicolumn{1}{c}{IRAS}&\multicolumn{1}{c}{IRAS}&\multicolumn{1}{c}{IRAS}&\multicolumn{1}{c}{IRAS}&
\multicolumn{1}{c}{MSX}&\multicolumn{1}{c}{AKARI}&\multicolumn{1}{c}{AKARI}&\multicolumn{1}{c}{AKARI}&
\multicolumn{1}{c}{AKARI}&\multicolumn{1}{c}{AKARI}&\multicolumn{1}{c}{AKARI}\\
  &   &   &   &   &
\multicolumn{1}{c}{[12]}&\multicolumn{1}{c}{[25]}&\multicolumn{1}{c}{[60]}&\multicolumn{1}{c}{[100]}&
\multicolumn{1}{c}{[8.3]}&\multicolumn{1}{c}{[9]}&\multicolumn{1}{c}{[18]}&\multicolumn{1}{c}{[65]}&
\multicolumn{1}{c}{[90]}&\multicolumn{1}{c}{[140]}&\multicolumn{1}{c}{[160]}\\
\noalign{\vspace{1mm}}
\hline
\noalign{\vspace{1mm}}
\endfirsthead
\multicolumn{16}{l}{{\smallbf\ \ Table 2.}{\small\ Continued\lstrut}}\\
\hline
SL&\multicolumn{1}{c}{$F$}&\multicolumn{1}{c}{$J$}&\multicolumn{1}{c}{$H$}&\multicolumn{1}{c}{$K$}\hstrut&
\multicolumn{1}{c}{IRAS}&\multicolumn{1}{c}{IRAS}&\multicolumn{1}{c}{IRAS}&\multicolumn{1}{c}{IRAS}&
\multicolumn{1}{c}{MSX}&\multicolumn{1}{c}{AKARI}&\multicolumn{1}{c}{AKARI}&\multicolumn{1}{c}{AKARI}&
\multicolumn{1}{c}{AKARI}&\multicolumn{1}{c}{AKARI}&\multicolumn{1}{c}{AKARI}\\
  &   &   &   &   &
\multicolumn{1}{c}{[12]}&\multicolumn{1}{c}{[25]}&\multicolumn{1}{c}{[60]}&\multicolumn{1}{c}{[100]}&
\multicolumn{1}{c}{[8.3]}&\multicolumn{1}{c}{[9]}&\multicolumn{1}{c}{[18]}&\multicolumn{1}{c}{[65]}&
\multicolumn{1}{c}{[90]}&\multicolumn{1}{c}{[140]}&\multicolumn{1}{c}{[160]}\\
\noalign{\vspace{1mm}}
\hline
\noalign{\vspace{1mm}}
\endhead
\endfoot
1	&-12.137	&-11.202	&-10.932	&-10.821	&-10.058	&-10.088	&-9.623	&-9.496	&-	&-9.986	 &-10.139	&-9.824	&-9.606	&-9.454	&-9.606	\\
3	&-12.277	&-11.394	&-10.928	&-10.657	&-	&-10.541	&-	&-	&-	&-10.282	&-10.426	&-	 &-10.625	&-	&-	\\
4	&-12.197	&-10.834	&-10.388	&-10.241	&-	&-10.523	&-10.678	&-	&-	&-10.419	&-10.390	 &-	&-	&-	&-	\\
5	&-11.853	&-11.194	&-10.792	&-10.729	&-	&-	&-10.561	&-	&-	&-	&-	&-	&-	&-	&-	\\
6	&-12.369	&-11.294	&-10.900	&-10.637	&-	&-	&-	&-	&-	&-10.207	&-	&-	&-	&-	&-	\\
7	&-12.933	&-11.458	&-11.068	&-10.857	&-	&-	&-	&-9.313	&-	&-	&-	&-	&-10.309	&-	&-	 \\
8	&-12.349	&-11.210	&-10.704	&-10.533	&-	&-10.115	&-10.171	&-9.653	&-	&-	&-	&-	 &-10.318	&-9.910	&-	\\
9	&-11.373	&-10.614	&-10.456	&-10.393	&-	&-10.319	&-9.771	&-	&-	&-10.121	&-10.311	 &-	&-9.763	&-9.353	&-9.765	\\
10	&-	&-11.162	&-10.660	&-10.465	&-9.688	&-9.804	&-8.963	&-	&-	&-10.080	&-10.203	&-9.321	 &-9.089	&-9.252	&-9.384	\\
11	&-12.721	&-11.214	&-10.752	&-10.561	&-10.097	&-9.921	&-9.456	&-	&-	&-	&-	&-	&-	&-	 &-	\\
13	&-12.461	&-11.318	&-11.068	&-10.989	&-	&-	&-	&-	&-	&-	&-	&-	&-	&-9.894	&-	\\
14	&-12.893	&-11.234	&-10.812	&-10.649	&-	&-	&-	&-	&-10.266	&-	&-	&-	&-	&-	&-	\\
15	&-12.729	&-10.986	&-10.544	&-10.489	&-	&-	&-	&-	&-	&-10.378	&-10.354	&-	&-	&-	 &-	\\
16	&-	&-11.570	&-10.952	&-10.577	&-9.658	&-9.548	&-	&-	&-	&-10.129	&-10.042	&-9.572	 &-9.594	&-9.310	&-9.404	\\
18	&-12.085	&-11.038	&-10.764	&-10.689	&-	&-	&-	&-	&-	&-	&-	&-8.449	&-8.843	&-8.377	 &-8.139	\\
21	&-10.985	&-10.442	&-10.156	&-10.037	&-	&-	&-	&-	&-	&-10.158	&-10.363	&-	&-	&-	 &-	\\
23	&-	&-10.562	&-10.128	&-9.897	&-	&-	&-	&-	&-	&-	&-	&-6.694	&-7.107	&-6.947	&-6.695	\\
25	&-12.645	&-11.114	&-10.800	&-10.701	&-	&-	&-	&-	&-	&-	&-	&-	&-8.487	&-8.485	&-8.266	 \\
27	&-	&-10.242	&-9.712	&-9.585	&-6.759	&-6.188	&-	&-5.899	&-	&-6.923	&-	&-6.748	&-	&-6.975	&-	\\
34	&-	&-9.866	&-9.020	&-8.741	&-9.305	&-	&-	&-	&-	&-9.391	&-	&-	&-	&-	&-	\\
35	&-	&-11.126	&-10.776	&-10.689	&-9.326	&-	&-	&-	&-	&-	&-	&-	&-	&-8.769	&-8.492	\\
36	&-	&-11.502	&-10.872	&-10.689	&-	&-	&-	&-	&-	&-	&-	&-8.476	&-8.750	&-8.491	&-8.447	\\
37	&-10.253	&-9.298	&-9.136	&-9.081	&-	&-	&-	&-	&-	&-8.963	&-	&-8.151	&-	&-8.666	&-	\\
38	&-	&-11.530	&-10.940	&-10.761	&-9.745	&-9.486	&-9.186	&-8.834	&-	&-10.447	&-9.825	&-9.433	 &-9.351	&-9.383	&-9.386	\\
39	&-12.773	&-11.110	&-10.740	&-10.685	&-	&-	&-	&-	&-	&-10.600	&-10.286	&-	&-9.570	 &-	&-	\\
40	&-	&-10.594	&-10.300	&-10.245	&-	&-	&-	&-	&-	&-9.420	&-	&-8.936	&-	&-	&-9.129	\\
41	&-12.301	&-11.194	&-10.880	&-10.797	&-9.839	&-9.986	&-9.871	&-9.402	&-	&-10.213	&-10.207	 &-	&-10.207	&-9.948	&-	\\
43	&-	&-11.126	&-10.724	&-10.669	&-8.650	&-	&-	&-7.442	&-	&-	&-	&-	&-	&-	&-	\\
44	&-	&-10.890	&-10.264	&-10.049	&-8.474	&-8.338	&-7.520	&-	&-	&-	&-	&-7.631	&-7.816	&-7.972	 &-7.951	\\
48	&-	&-10.614	&-10.272	&-10.149	&-9.206	&-9.423	&-8.689	&-8.404	&-	&-9.211	&-9.819	&-8.684	 &-8.750	&-8.752	&-8.792	\\
49	&-11.477	&-10.782	&-10.364	&-10.153	&-9.498	&-	&-	&-	&-	&-9.518	&-9.745	&-9.175	&-9.161	 &-9.387	&-9.320	\\
50	&-12.645	&-11.270	&-10.592	&-10.205	&-9.727	&-9.018	&-	&-	&-	&-9.709	&-9.582	&-9.798	 &-9.896	&-9.731	&-	\\
51	&-12.421	&-11.314	&-10.968	&-10.897	&-	&-	&-	&-	&-	&-	&-	&-	&-10.076	&-9.723	&-	 \\
52	&-	&-11.102	&-9.960	&-9.401	&-8.835	&-8.883	&-8.687	&-8.745	&-	&-8.831	&-8.954	&-8.635	&-8.880	 &-8.980	&-8.904	\\
54	&-11.809	&-9.870	&-9.276	&-8.945	&-8.888	&-8.533	&-8.075	&-	&-	&-8.864	&-8.728	&-	&-	&-	&-	\\
56	&-	&-10.518	&-9.372	&-8.669	&-8.206	&-8.110	&-8.051	&-	&-	&-8.096	&-8.161	&-8.358	&-8.497	&-8.436	 &-8.162	\\
57	&-12.869	&-11.222	&-10.920	&-10.845	&-9.870	&-	&-	&-	&-	&-	&-	&-	&-	&-	&-	\\
58	&-12.057	&-11.174	&-10.776	&-10.673	&-	&-	&-	&-	&-	&-10.520	&-	&-	&-	&-	&-	\\
60	&-11.717	&-10.618	&-10.304	&-10.229	&-9.854	&-	&-	&-	&-	&-10.180	&-10.063	&-9.467	 &-9.451	&-9.319	&-9.343	\\
61	&-	&-11.170	&-10.460	&-10.141	&-	&-	&-	&-	&-	&-9.804	&-9.935	&-9.398	&-9.278	&-9.184	 &-9.284	\\
62	&-12.557	&-11.046	&-10.820	&-10.761	&-9.721	&-9.444	&-9.824	&-	&-	&-10.054	&-9.857	&-	 &-9.820	&-	&-	\\
63	&-11.493	&-10.414	&-10.132	&-10.061	&-	&-9.565	&-	&-	&-	&-	&-	&-	&-	&-	&-	\\
64	&-12.945	&-10.938	&-10.512	&-10.385	&-8.742	&-9.013	&-8.102	&-	&-	&-	&-	&-	&-8.729	&-	 &-	\\
65	&-12.381	&-11.314	&-10.872	&-10.685	&-	&-10.365	&-	&-8.811	&-	&-10.286	&-10.380	 &-9.960	&-9.751	&-	&-9.272	\\
66	&-	&-11.578	&-10.864	&-10.601	&-8.303	&-7.594	&-7.416	&-7.489	&-	&-	&-	&-	&-	&-7.675	 &-7.601	\\
68	&-	&-11.026	&-10.940	&-10.849	&-	&-	&-	&-9.874	&-	&-	&-	&-	&-	&-	&-	\\
71	&-	&-11.110	&-10.916	&-10.789	&-	&-	&-	&-	&-	&-9.049	&-	&-	&-	&-	&-	\\
72	&-	&-11.138	&-10.372	&-10.145	&-8.429	&-8.008	&-7.264	&-7.201	&-	&-	&-	&-	&-	&-	&-	\\
74	&-12.501	&-11.230	&-10.832	&-10.745	&-	&-	&-	&-	&-	&-	&-10.619	&-	&-10.554	 &-9.789	&-	\\
75	&-10.601	&-10.086	&-9.720	&-9.581	&-9.553	&-9.680	&-	&-	&-	&-9.615	&-9.749	&-	&-	&-	&-	\\
77	&-11.625	&-11.406	&-11.056	&-10.893	&-	&-10.402	&-9.602	&-	&-	&-	&-	&-	&-10.167	 &-10.046	&-	\\
78	&-10.169	&-9.878	&-9.728	&-9.657	&-	&-	&-	&-	&-10.163	&-10.124	&-10.662	&-	&-10.724	 &-10.348	&-	\\
79	&-10.977	&-10.546	&-10.396	&-10.309	&-10.187	&-10.287	&-	&-	&-	&-10.187	&-10.564	 &-	&-	&-	&-	\\
80	&-	&-10.522	&-10.048	&-9.857	&-8.111	&-7.323	&-7.276	&-7.409	&-	&-8.132	&-7.448	&-7.256	&-	 &-8.056	&-8.019	\\
81	&-12.897	&-10.994	&-10.544	&-10.425	&-10.011	&-10.188	&-	&-9.077	&-10.266	&-10.040	 &-10.168	&-	&-	&-	&-	\\
82	&-8.653	&-9.274	&-9.024	&-8.957	&-9.321	&-9.599	&-9.331	&-8.953	&-9.315	&-9.350	&-9.628	&-9.438	&-9.288	 &-9.345	&-9.401	\\
83	&-12.553	&-10.806	&-10.196	&-9.929	&-9.939	&-9.800	&-9.590	&-	&-	&-9.827	&-9.824	&-	&-9.898	 &-9.746	&-	\\
84	&-11.813	&-11.326	&-11.084	&-11.001	&-	&-	&-	&-	&-	&-	&-	&-	&-9.801	&-9.991	&-9.981	 \\
85	&-10.665	&-10.602	&-10.392	&-10.329	&-	&-	&-	&-	&-	&-10.413	&-	&-	&-	&-	&-	\\
86	&-12.493	&-10.470	&-10.008	&-9.905	&-	&-10.268	&-10.310	&-	&-	&-10.338	&-10.389	 &-	&-10.300	&-9.711	&-9.996	\\
87	&-	&-10.570	&-10.308	&-10.261	&-9.516	&-9.908	&-8.933	&-8.476	&-	&-9.271	&-	&-8.885	&-8.995	 &-8.963	&-9.118	\\
88	&-11.897	&-10.930	&-10.608	&-10.505	&-9.290	&-9.211	&-9.354	&-9.206	&-	&-	&-	&-9.586	 &-9.589	&-9.506	&-	\\
89	&-12.593	&-10.838	&-10.460	&-10.409	&-10.222	&-9.896	&-	&-	&-	&-10.346	&-9.966	&-	 &-9.798	&-9.379	&-	\\
93	&-	&-11.442	&-10.904	&-10.793	&-	&-9.853	&-	&-	&-	&-10.451	&-	&-	&-	&-	&-	\\
95	&-11.997	&-9.798	&-8.928	&-8.337	&-7.645	&-7.458	&-7.447	&-7.460	&-7.704	&-7.785	&-7.558	&-	&-	&-	 &-	\\
96	&-	&-11.394	&-10.892	&-10.753	&-	&-	&-	&-	&-	&-	&-10.729	&-	&-9.955	&-	&-	\\
98	&-12.133	&-10.698	&-10.448	&-10.357	&-	&-	&-	&-	&-	&-10.254	&-	&-	&-	&-	&-	\\
100	&-12.045	&-10.906	&-10.520	&-10.413	&-	&-	&-	&-	&-	&-	&-	&-	&-10.399	&-9.761	 &-9.614	\\
101	&-10.889	&-10.102	&-10.004	&-9.957	&-	&-	&-	&-	&-10.401	&-10.233	&-	&-	&-	&-	&-	 \\
102	&-12.449	&-10.958	&-10.640	&-10.553	&-	&-10.690	&-	&-	&-	&-	&-10.448	&-	&-10.370	 &-9.910	&-	\\
104	&-12.585	&-11.298	&-11.004	&-10.897	&-	&-9.967	&-10.301	&-9.350	&-inf	&-	&-	&-	&-	 &-	&-	\\
106	&-11.177	&-10.774	&-10.396	&-10.273	&-10.125	&-10.222	&-	&-	&-	&-9.844	&-10.158	 &-	&-	&-	&-	\\
107	&-13.113	&-11.502	&-11.176	&-11.085	&-	&-	&-10.462	&-	&-	&-	&-	&-	&-	&-	&-	\\
108	&-11.985	&-11.098	&-10.704	&-10.525	&-8.499	&-8.286	&-7.605	&-7.591	&-	&-	&-	&-	&-	&-	 &-	\\
109	&-11.777	&-10.682	&-10.324	&-10.149	&-9.398	&-9.283	&-9.218	&-9.173	&-	&-9.430	&-9.345	 &-9.320	&-9.259	&-9.417	&-9.488	\\
110	&-12.313	&-11.270	&-11.056	&-10.937	&-	&-	&-10.502	&-	&-	&-	&-	&-	&-10.721	&-	 &-	\\
111	&-12.365	&-10.918	&-10.512	&-10.365	&-	&-10.377	&-10.620	&-9.743	&-	&-10.358	 &-10.403	&-	&-	&-	&-	\\
112	&-11.109	&-10.434	&-10.336	&-10.269	&-10.155	&-	&-	&-	&-10.266	&-10.238	&-10.565	 &-	&-	&-	&-	\\
113	&-11.785	&-10.894	&-10.708	&-10.641	&-	&-10.429	&-10.638	&-	&-	&-	&-	&-	&-10.544	 &-	&-	\\
114	&-11.929	&-11.002	&-10.872	&-10.813	&-	&-	&-	&-	&-	&-	&-	&-	&-10.132	&-	&-9.923	 \\
115	&-11.869	&-10.986	&-10.688	&-10.377	&-	&-10.150	&-	&-9.093	&-	&-	&-10.155	&-	&-	 &-	&-	\\
116	&-11.905	&-10.438	&-10.064	&-9.909	&-9.606	&-9.739	&-	&-	&-	&-	&-9.899	&-	&-9.730	&-9.523	 &-9.633	\\
118	&-	&-11.318	&-10.680	&-10.369	&-	&-10.180	&-9.996	&-	&-10.044	&-10.101	&-10.288	 &-	&-10.151	&-9.880	&-9.847	\\
120	&-	&-11.510	&-11.036	&-10.809	&-10.034	&-10.240	&-10.225	&-	&-	&-10.254	&-9.982	 &-	&-	&-9.804	&-9.748	\\
121	&-11.933	&-10.638	&-10.348	&-10.261	&-	&-	&-	&-	&-	&-10.338	&-	&-	&-	&-	&-	\\
123	&-	&-9.826	&-9.108	&-8.913	&-	&-	&-	&-	&-	&-9.753	&-	&-	&-	&-	&-	\\
124	&-11.025	&-10.098	&-9.824	&-9.681	&-9.357	&-9.290	&-9.233	&-	&-	&-9.395	&-9.406	&-9.346	&-9.454	 &-	&-	\\
125	&-12.145	&-11.122	&-10.716	&-10.621	&-	&-10.402	&-10.063	&-9.947	&-	&-	&-	&-	 &-10.365	&-	&-	\\
126	&-11.333	&-11.234	&-11.008	&-10.945	&-	&-10.319	&-10.128	&-	&-	&-	&-10.438	&-	 &-10.159	&-9.679	&-9.775	\\
129	&-	&-11.486	&-11.180	&-11.073	&-8.393	&-	&-	&-7.088	&-	&-	&-	&-	&-	&-	&-	\\
130	&-11.261	&-9.786	&-9.256	&-8.897	&-8.197	&-8.285	&-8.380	&-8.289	&-	&-8.403	&-8.440	&-8.387	&-8.654	 &-8.837	&-8.880	\\
131	&-10.529	&-9.938	&-9.604	&-9.465	&-9.317	&-9.460	&-9.561	&-9.239	&-9.328	&-9.347	&-9.426	&-	&-	&-	 &-	\\
132	&-11.741	&-9.218	&-8.676	&-8.589	&-	&-	&-	&-	&-9.679	&-9.739	&-10.532	&-	&-	&-	&-	\\
133	&-12.797	&-11.534	&-11.100	&-11.017	&-	&-	&-10.393	&-9.964	&-	&-	&-	&-	&-10.385	 &-	&-	\\
134	&-12.313	&-11.186	&-10.836	&-10.753	&-	&-	&-10.545	&-	&-	&-	&-	&-	&-10.772	&-	 &-	\\
135	&-12.501	&-11.370	&-11.172	&-11.113	&-10.155	&-10.240	&-9.898	&-	&-	&-	&-	&-	 &-9.916	&-9.802	&-9.620	\\
137	&-12.341	&-11.178	&-10.912	&-10.793	&-	&-10.095	&-10.000	&-	&-	&-	&-10.480	&-	 &-10.191	&-	&-	\\
138	&-12.013	&-11.166	&-10.976	&-10.921	&-10.058	&-9.967	&-10.041	&-	&-	&-	&-	&-	 &-9.915	&-9.626	&-9.595	\\
140	&-11.121	&-10.702	&-10.524	&-10.461	&-10.022	&-9.769	&-9.220	&-9.005	&-	&-	&-	&-9.366	 &-	&-9.230	&-9.247	\\
141	&-12.177	&-11.546	&-11.228	&-11.169	&-9.939	&-9.444	&-	&-	&-	&-	&-	&-	&-	&-	&-	\\
142	&-11.693	&-10.086	&-9.500	&-9.197	&-9.488	&-	&-	&-	&-	&-9.358	&-10.125	&-	&-	&-	&-	\\
143	&-	&-11.650	&-11.380	&-11.417	&-	&-	&-	&-	&-	&-	&-	&-	&-9.825	&-9.461	&-	\\
148	&-12.593	&-11.242	&-10.972	&-11.001	&-	&-9.853	&-	&-	&-	&-	&-	&-8.798	&-9.023	&-8.736	 &-8.895	\\
155	&-12.353	&-11.322	&-11.128	&-11.181	&-	&-10.082	&-9.824	&-	&-	&-	&-	&-	&-	&-	&-	 \\
158	&-10.781	&-10.054	&-9.844	&-9.801	&-10.084	&-10.416	&-	&-	&-10.044	&-10.020	&-10.486	 &-	&-10.634	&-	&-9.707	\\
159	&-12.169	&-11.042	&-10.820	&-10.857	&-	&-10.523	&-	&-	&-	&-	&-	&-	&-	&-	&-	\\
160	&-13.069	&-11.518	&-11.316	&-11.349	&-	&-	&-	&-	&-	&-	&-	&-	&-10.789	&-	&-	\\
162	&-12.041	&-10.686	&-10.440	&-10.417	&-	&-	&-	&-	&-	&-	&-	&-	&-10.308	&-9.845	&-	 \\
163	&-11.933	&-10.974	&-10.820	&-10.777	&-	&-	&-10.190	&-	&-	&-	&-	&-	&-	&-	&-	\\
165	&-11.089	&-10.282	&-10.092	&-10.089	&-	&-	&-	&-	&-10.296	&-10.297	&-10.564	&-	 &-	&-	&-	\\
169	&-12.417	&-11.258	&-10.960	&-10.997	&-10.000	&-10.150	&-9.500	&-9.094	&-	&-	&-	&-	 &-	&-	&-	\\
171	&-11.569	&-9.430	&-9.028	&-8.997	&-	&-	&-	&-	&-9.851	&-9.960	&-	&-	&-	&-	&-	\\
172	&-12.377	&-11.318	&-11.152	&-11.193	&-10.140	&-9.143	&-9.347	&-	&-	&-	&-	&-	&-	&-	 &-	\\
174	&-12.789	&-11.554	&-11.380	&-11.397	&-	&-	&-10.240	&-9.601	&-	&-	&-	&-	&-	&-	&-	 \\
175	&-10.301	&-10.106	&-10.188	&-10.329	&-9.727	&-10.389	&-10.122	&-	&-	&-10.480	&-	 &-	&-10.559	&-	&-	\\
176	&-10.725	&-10.378	&-10.360	&-10.465	&-	&-10.620	&-	&-	&-	&-	&-10.604	&-	&-	&-	 &-	\\
177	&-11.481	&-10.682	&-10.548	&-10.625	&-	&-	&-9.528	&-	&-	&-	&-	&-	&-	&-	&-	\\
178	&-12.893	&-11.298	&-11.156	&-11.257	&-	&-	&-	&-9.317	&-	&-	&-	&-	&-	&-	&-	\\
179	&-	&-11.570	&-11.480	&-11.589	&-	&-	&-10.382	&-	&-	&-	&-	&-	&-	&-	&-	\\
180	&-10.953	&-10.238	&-10.172	&-10.229	&-	&-10.389	&-	&-	&-	&-10.318	&-10.357	&-	 &-	&-	&-	\\
181	&-12.837	&-11.506	&-11.424	&-11.533	&-	&-	&-10.190	&-	&-	&-	&-	&-	&-	&-9.740	&-	 \\
182	&-13.069	&-11.538	&-11.420	&-11.521	&-	&-	&-10.545	&-9.563	&-	&-	&-	&-	&-10.666	 &-	&-	\\
183	&-10.537	&-10.362	&-10.420	&-10.537	&-10.111	&-9.939	&-	&-	&-	&-10.116	&-10.028	 &-	&-10.775	&-	&-	\\
184	&-11.841	&-10.626	&-10.416	&-10.493	&-	&-10.523	&-	&-	&-	&-10.537	&-10.838	&-	 &-	&-	&-	\\
185	&-13.129	&-11.438	&-11.292	&-11.389	&-	&-	&-	&-9.596	&-	&-	&-	&-	&-10.708	&-9.986	 &-	\\
186	&-12.373	&-11.238	&-11.048	&-11.125	&-	&-10.287	&-	&-9.498	&-	&-	&-	&-	&-	&-	&-	 \\
187	&-12.261	&-11.078	&-10.996	&-11.149	&-	&-	&-10.530	&-	&-	&-	&-	&-	&-	&-	&-	\\
\noalign{\vspace{1mm}}
\hline

\end{longtable}

\vbox{
\tabcolsep=2pt
\begin{longtable}{lD..{3.3}D..{3.3}D..{3.3}D..{3.3}D..{3.3}D..{3.3}D..{3.3}D..{3.3}
D..{3.3}D..{3.3}D..{3.3}D..{3.3}D..{3.3}D..{3.3}D..{3.3}}
\multicolumn{16}{l}{{\smallbf\ \ Table 2.}{\small\ Continued\lstrut}}\\
\hline
\hhuad ID&
\multicolumn{1}{c}{$F$}&\multicolumn{1}{c}{$J$}&\multicolumn{1}{c}{$H$}&\multicolumn{1}{c}{$K$}\hstrut&
\multicolumn{1}{c}{IRAS}&\multicolumn{1}{c}{IRAS}&\multicolumn{1}{c}{IRAS}&\multicolumn{1}{c}{IRAS}&
\multicolumn{1}{c}{MSX}&\multicolumn{1}{c}{AKARI}&\multicolumn{1}{c}{AKARI}&\multicolumn{1}{c}{AKARI}&
\multicolumn{1}{c}{AKARI}&\multicolumn{1}{c}{AKARI}&\multicolumn{1}{c}{AKARI}\\
  &   &   &   &   &
\multicolumn{1}{c}{[12]}&\multicolumn{1}{c}{[25]}&\multicolumn{1}{c}{[60]}&\multicolumn{1}{c}{[100]}&
\multicolumn{1}{c}{[8.3]}&\multicolumn{1}{c}{[9]}&\multicolumn{1}{c}{[18]}&\multicolumn{1}{c}{[65]}&
\multicolumn{1}{c}{[90]}&\multicolumn{1}{c}{[140]}&\multicolumn{1}{c}{[160]}\\
\noalign{\vspace{1mm}}
\hline
KW97\,14-24	&-10.289	&-10.133	&-10.018	&-9.997	&-8.303	&-7.595	&-7.416	&-7.488	&-9.910	&-9.000	&-	 &-	&-	&-\hstrut&-	\\
KW97\,14-52	&-10.369	&-8.986	&-8.756	&-8.913	&-	&-	&-	&-	&-	&-10.304	&-	&-	&-	&-	&-	\\
IRAS\,03243+5819	&-11.229	&-10.069	&-9.965	&-9.819	&-9.770	&-10.620	&-	&-	&-9.726	&-9.176	 &-10.349	&-	&-	&-	&-	\\
LkH$alpha$\,272	&-10.249	&-9.580	&-9.489	&-9.577	&-	&-10.213	&-10.000	&-9.620	&-	&-10.199	 &-10.264	&-	&-	&-	&-	\\
XY\,Per	&-8.809	&-8.480	&-8.462	&-8.486	&-9.017	&-9.312	&-9.610	&-9.653	&-	&-8.843	&-9.246	&-9.738	&-9.837	 &-9.943	&-	\\
Gahm\,21	&-10.433	&-10.321	&-10.160	&-10.237	&-	&-	&-	&-9.155	&-	&-	&-	&-	&-	&-	&-	 \\
KW9716\,55	&-10.289	&-9.875	&-9.901	&-10.056	&-	&-10.506	&-9.602	&-	&-	&-	&-	&-	&-	&-	&-	 \\
OS\,Per	&-	&-9.594	&-9.605	&-9.829	&-10.125	&-9.444	&-	&-	&-	&-	&-	&-	&-	&-	&-	\\
V347\,Aur	&-10.405	&-9.378	&-9.225	&-9.273	&-9.385	&-9.313	&-9.342	&-9.324	&-	&-9.548	&-9.519	&-9.569	 &-9.578	&-9.731	&-9.897	\\
\noalign{\vspace{1mm}}
\hline
\end{longtable}
}
\endlandscape

\noindent
\tabcolsep=3pt
\vbox{
\footnotesize
\centerline{\parbox{110mm}{\baselineskip=9pt
{\smallbf\ \ Table 3.}{\small\ The values of
$\log \lambda F_{\lambda}$ for the suspected YSOs
in passbands of the GSC~2.3.2, 2MASS and Spitzer surveys.}}}
\vskip-2mm

\begin{longtable}{rrrrrrrrrr}
\hline
SL & $F$~~~~ & $J$~~~~ & $H$~~~ & $K_s$~~ &  Spitzer  & Spitzer  &
Spitzer & Spitzer  & Spitzer\hstrut  \\
    & [0.71]~ & [1.26]~  & [1.6]~  &  [2.2]~~  &   [3.6]~~  &  [4.5]~~
&   [5.8]~~  &   [8.0]~~  &   [24.0]~\lstrut \\
\hline
 18  &  -12.085 &  -11.038 &  -10.764&   -10.689 &  -10.679 &  -10.801 &  -11.006 &  -11.308 &  -12.061\hstrut  \\
 20  &    --~~~~ &  -11.410 &  -10.764&   -10.413 &  -10.235 &  -10.277 &  -10.286 &  -10.292 &  -10.221  \\
 30  &    --~~~~ &  -11.458 &  -11.036&   -10.945 &  -10.903 &  -10.977 &  -11.190 &  -11.264 & --~~~~  \\
 31  &    --~~~~ &  -11.238 &  -10.812&   -10.589 &  -10.355 &  -10.397 &  -10.530 &  -10.552 &  -10.665  \\
 37  &  -10.253 &   -9.298 &   -9.136&    -9.081 &  -10.239 &  -10.329 &  -10.638 &  -10.896 &  -11.633  \\
 51  &  -12.421 &  -11.314 &  -10.968&   -10.897 &  -10.915 &  -11.009 &  -11.026 &  -11.084 &  -11.301  \\
 52  &    --~~~~ &  -11.102 &   -9.960&    -9.401 &   -9.115 &   -9.077 &   -9.062 &    --~~~~ &    --~~~~  \\
 53  &  -12.293 &  -11.286 &  -11.032&   -10.937 &  -10.931 &  -11.033 &  -11.126 &  -11.092 &  -11.425  \\
 54  &  -11.809 &   -9.870 &   -9.276&    -8.945 &   -8.851 &   -8.853 &   -8.830 &   -8.856 &    --~~~~  \\
 55  &  -11.229 &  -10.682 &  -10.540&   -10.481 &  -10.611 &  -10.741 &  -10.846 &  -10.864 &  -10.741  \\
 56  &    --~~~~ &  -10.518 &   -9.372&    -8.669 &   -8.763 &   -8.673 &   -8.082 &    --~~~~ &    --~~~~  \\
 57  &  -12.869 &  -11.222 &  -10.920&   -10.845 &  -11.059 &  -11.041 &  -11.054 &  -11.004 &  -11.337  \\
 59  &  -11.957 &  -11.262 &  -10.952&   -10.889 &  -11.019 &  -11.141 &  -11.310 &  -11.444 &  -11.597  \\
 60  &  -11.717 &  -10.618 &  -10.304&   -10.229 &  -10.191 &  -10.225 &  -10.234 &  -10.252 &  -10.513  \\
 61  &    --~~~~ &  -11.170 &  -10.460&   -10.141 &  -10.039 &  -10.021 &  -10.078 &  -10.096 &  -10.289  \\
 62  &  -12.557 &  -11.046 &  -10.820&   -10.761 &  -10.571 &  -10.493 &  -10.442 &  -10.268 &  -10.041  \\
 64  &  -12.945 &  -10.938 &  -10.512&   -10.385 &  -10.447 &  -10.565 &  -10.646 &  -10.592 &  -10.881  \\
 65  &  -12.381 &  -11.314 &  -10.872&   -10.685 &  -10.591 &  -10.573 &  -10.670 &  -10.644 &  -10.689  \\
 66  &    --~~~~ &  -11.578 &  -10.864&   -10.601 &  -10.511 &  -10.509 &    --~~~~ &    --~~~~ &    --~~~~  \\
 69  &    --~~~~ &  -10.918 &  -10.432&   -10.209 &  -10.235 &  -10.213 &  -10.170 &  -10.000 &    --~~~~  \\
 70  &    --~~~~ &  -11.026 &  -10.480&   -10.289 &  -10.223 &  -10.269 &  -10.374 &  -10.496 &    --~~~~  \\
 71  &    --~~~~ &  -11.110 &  -10.916&   -10.789 &  -10.519 &  -10.453 &  -10.382 &  -10.080 &    --~~~~  \\
 72  &    --~~~~ &  -11.138 &  -10.372&   -10.145 &  -10.315 &  -10.433 &  -10.578 &  -10.528 &    --~~~~  \\
 74  &  -12.501 &  -11.230 &  -10.832&   -10.745 &  -10.507 &  -10.473 &  -10.494 &  -10.516 &    --~~~~  \\
 75  &  -10.601 &  -10.086 &   -9.720&    -9.581 &   -9.535 &   -9.601 &   -9.706 &   -9.720 &    --~~~~  \\
 80  &    --~~~~ &  -10.522 &  -10.048&    -9.857 &   -9.319 &   -9.105 &   -8.870 &   -8.692 &    --~~~~  \\
 89  &  -12.593 &  -10.838 &  -10.460&   -10.409 &  -10.447 &  -10.389 &  -10.370 &  -10.316 &   -9.981  \\
 91  &    --~~~~ &  -11.538 &  -11.156&   -11.081 &  -11.131 &  -11.229 &  -11.406 &  -11.624 &  -12.085  \\
 93  &    --~~~~ &  -11.442 &  -10.904&   -10.793 &  -10.891 &  -10.969 &  -11.066 &  -11.132 &  -11.213  \\
 94  &  -12.225 &  -10.774 &  -10.340&   -10.161 &   -9.971 &   -9.977 &  -10.062 &  -10.148 &  -10.325  \\
 96  &    --~~~~ &  -11.394 &  -10.892&   -10.753 &  -10.611 &  -10.649 &  -10.742 &  -10.784 &  -10.761  \\
 97  &    --~~~~ &  -11.330 &  -10.904&   -10.849 &  -10.891 &  -11.005 &  -11.178 &  -11.372 &  -11.805  \\
 145 &    --~~~~ &  -11.514 &  -11.188&   -11.185 &  -11.387 &  -11.505 &  -11.682 &  -11.896 &  -12.309  \\
 146 &  -12.445 &  -10.970 &  -10.692&   -10.725 &  -10.855 &  -11.009 &  -11.170 &  -11.224 &  -11.245  \\
 147 &    --~~~~ &  -11.290 &  -10.928&   -10.929 &  -11.035 &  -11.105 &  -11.210 &  -11.296 &  -11.845  \\
 148 &  -12.593 &  -11.242 &  -10.972&   -11.001 &  -11.107 &  -11.241 &  -11.374 &  -11.380 &  -11.469  \\
 149 &    --~~~~ &  -11.582 &  -11.260&   -11.285 &  -11.523 &  -11.625 &  -11.798 &  -11.976 &    --~~~~  \\
\noalign{\vspace{1mm}}
\hline
\end{longtable}
}

However, SEDs of dusty spiral galaxies in the mid- and far-infrared are
quite similar to SEDs of Stage I YSOs (Devriendt et al. 1999; Dale \&
Helou 2002; Sajina et al. 2006).  This is especially true for heavily
reddened and redshifted galaxies.  Their identification applying the
SEDs with only a few infrared photometric points is quite problematic --
high resolution optical imaging and infrared spectroscopy are essential
(see Luhmann et al. 2006; Rebull et al. 2010).

A few galaxies in our region, identified via the Simbad database, were
excluded during the compilation of lists of possible YSOs in Papers II
an III.  Now we checked for galaxies and extended infrared objects in
the 2MASX Database (Skrutskie et al. 2006), NASA/IPAC Extragalactic
Database NED (2010) and the Sloan Digital Sky Survey (SDSS) 7th data
release (Abazajian et al. 2009).  Also, the objects

\vbox{
\centerline{\psfig{figure=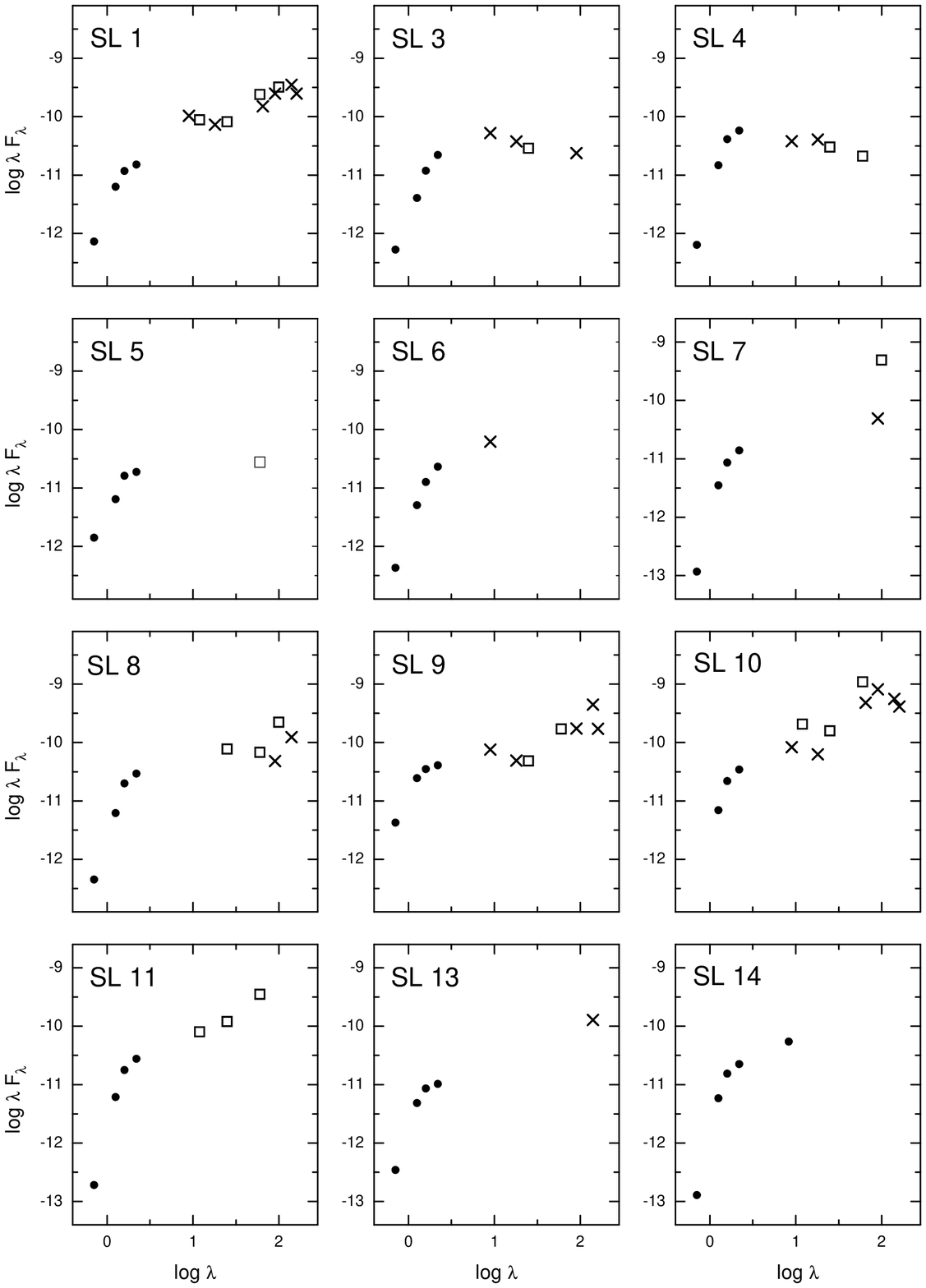,width=124mm,clip=}}
\captionb{1}{Spectral energy distributions of
the suspected YSOs.  Dots represent the results for $R$, $J$, $H$, $K_s$
and MSX passbands, circles are for the Spitzer passbands, squares for
the IRAS passbands and crosses for the AKARI passbands.}
}
\newpage

\vbox{
\centerline{\psfig{figure=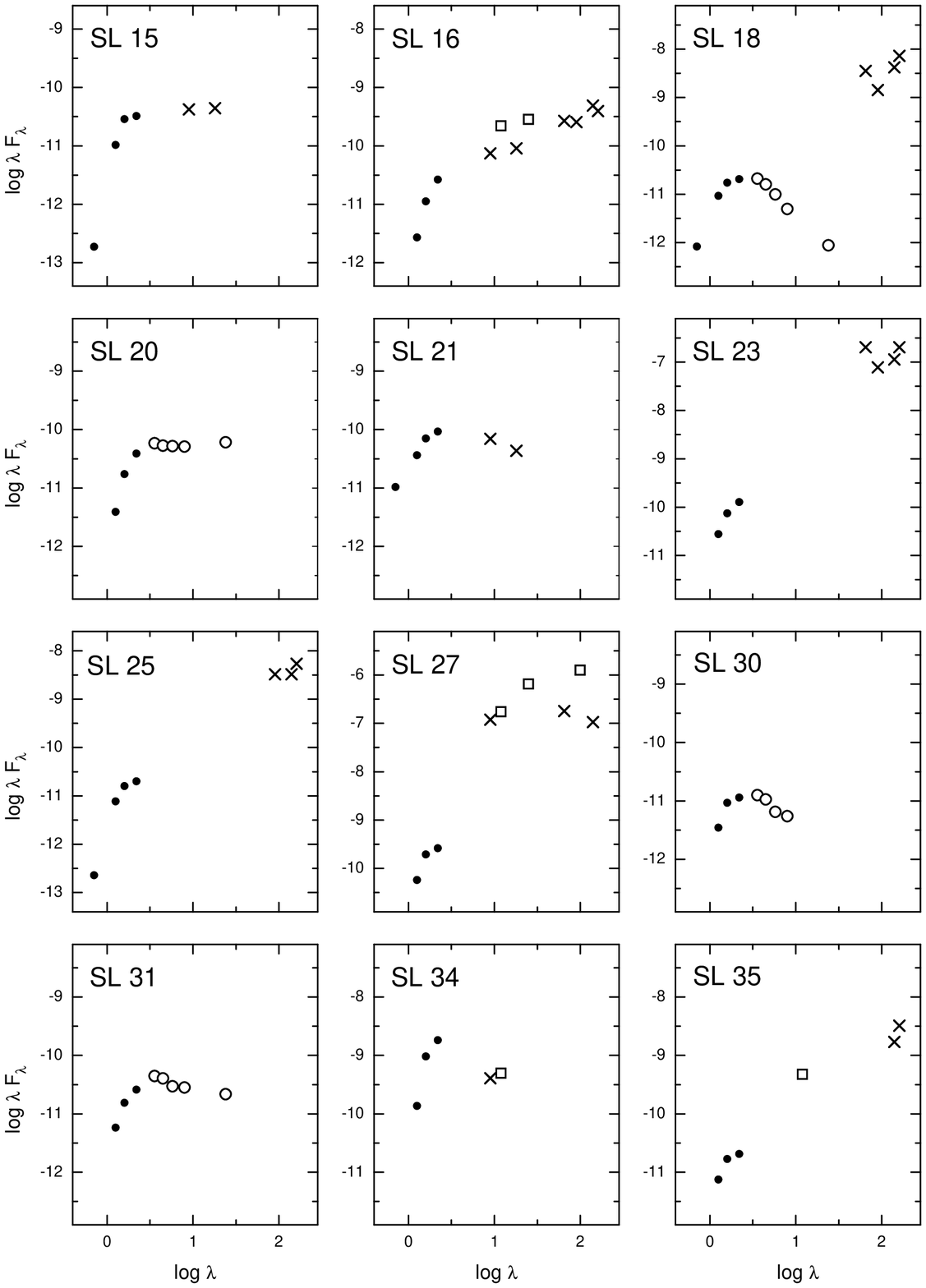,width=124mm,clip=}}
\captionc{1}{Continued}
}
\newpage

\vbox{
\centerline{\psfig{figure=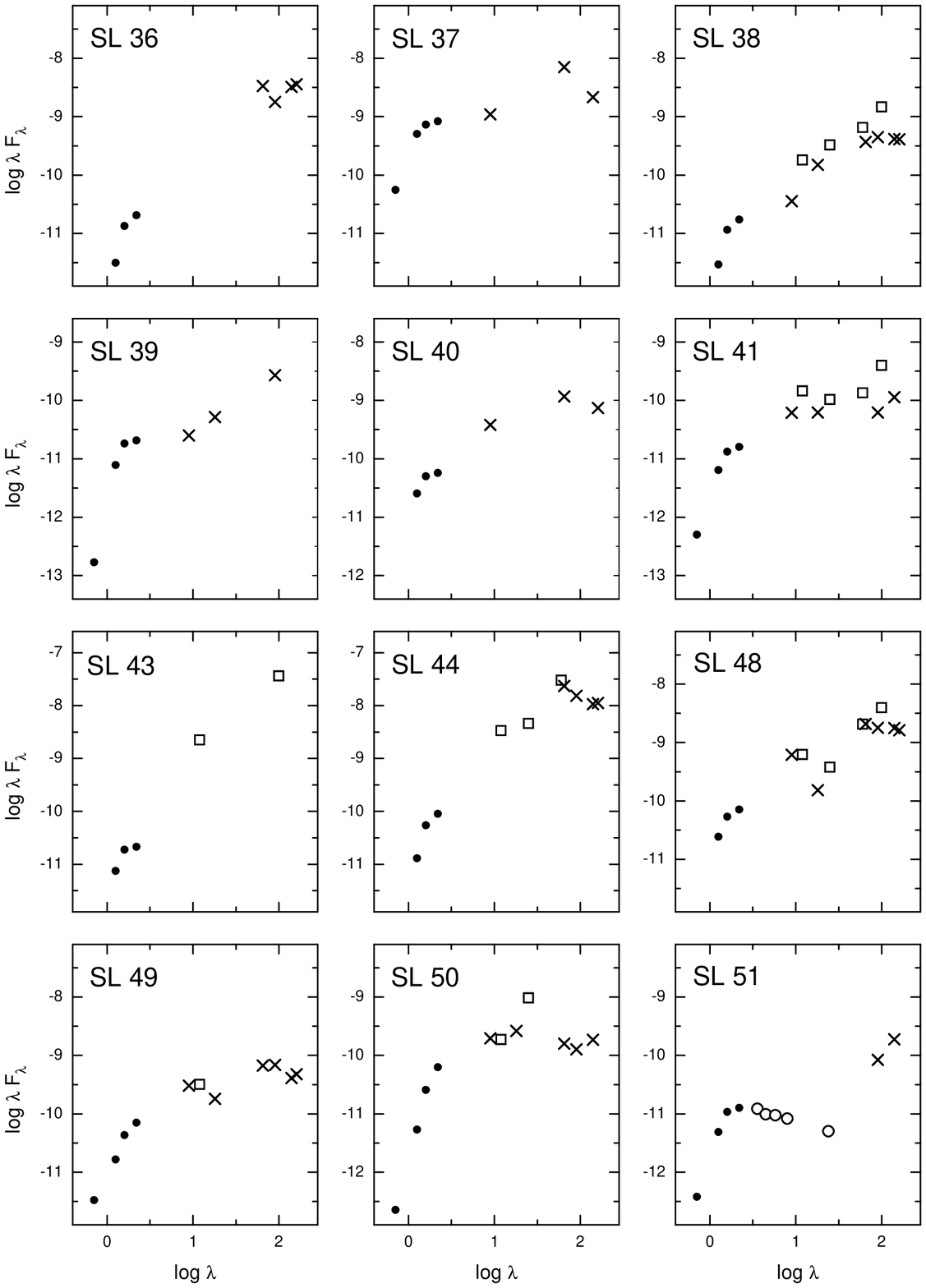,width=124mm,clip=}}
\captionc{1}{Continued}
}
\newpage

\vbox{
\centerline{\psfig{figure=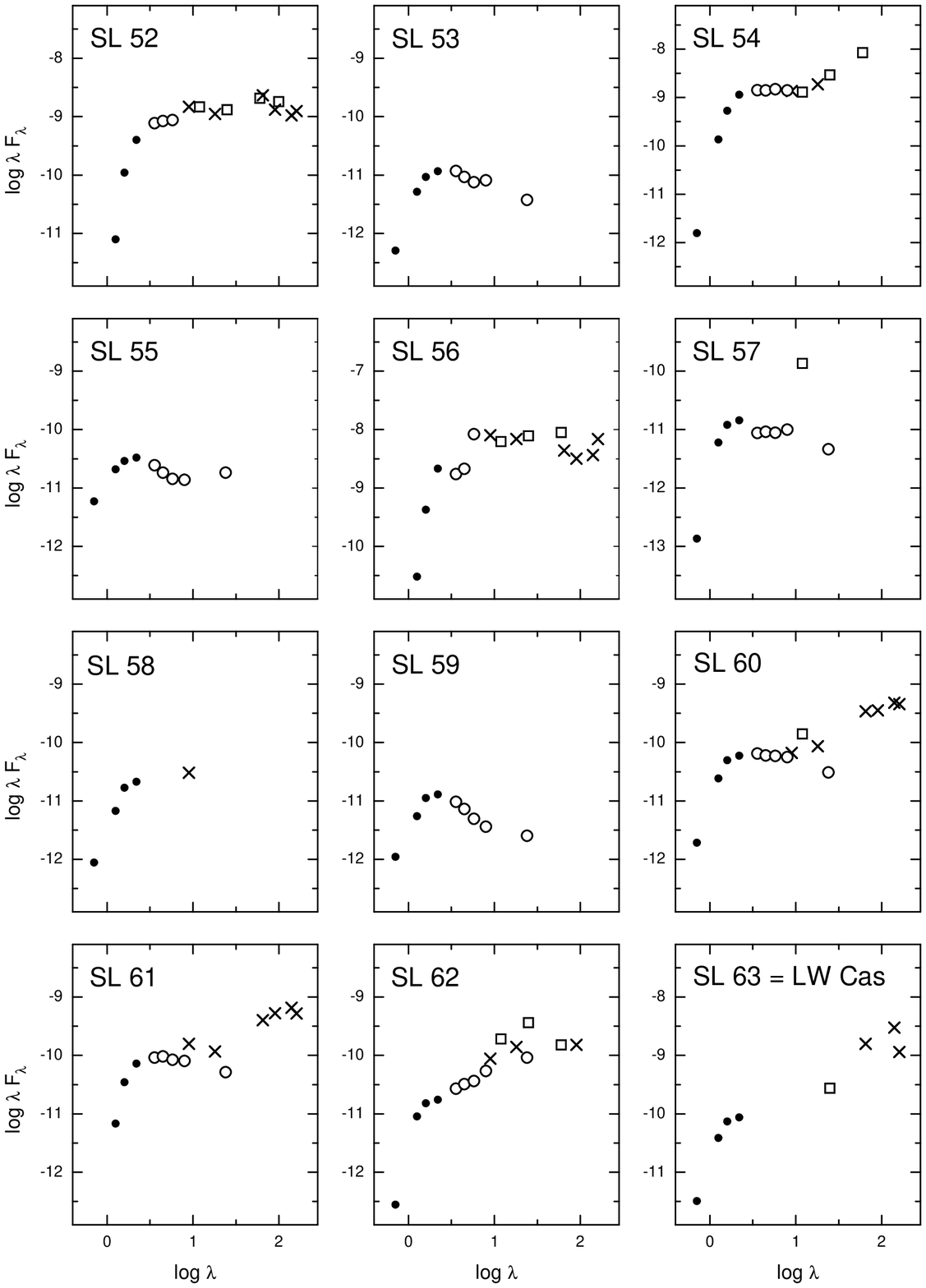,width=124mm,clip=}}
\captionc{1}{Continued}
}
\newpage

\vbox{
\centerline{\psfig{figure=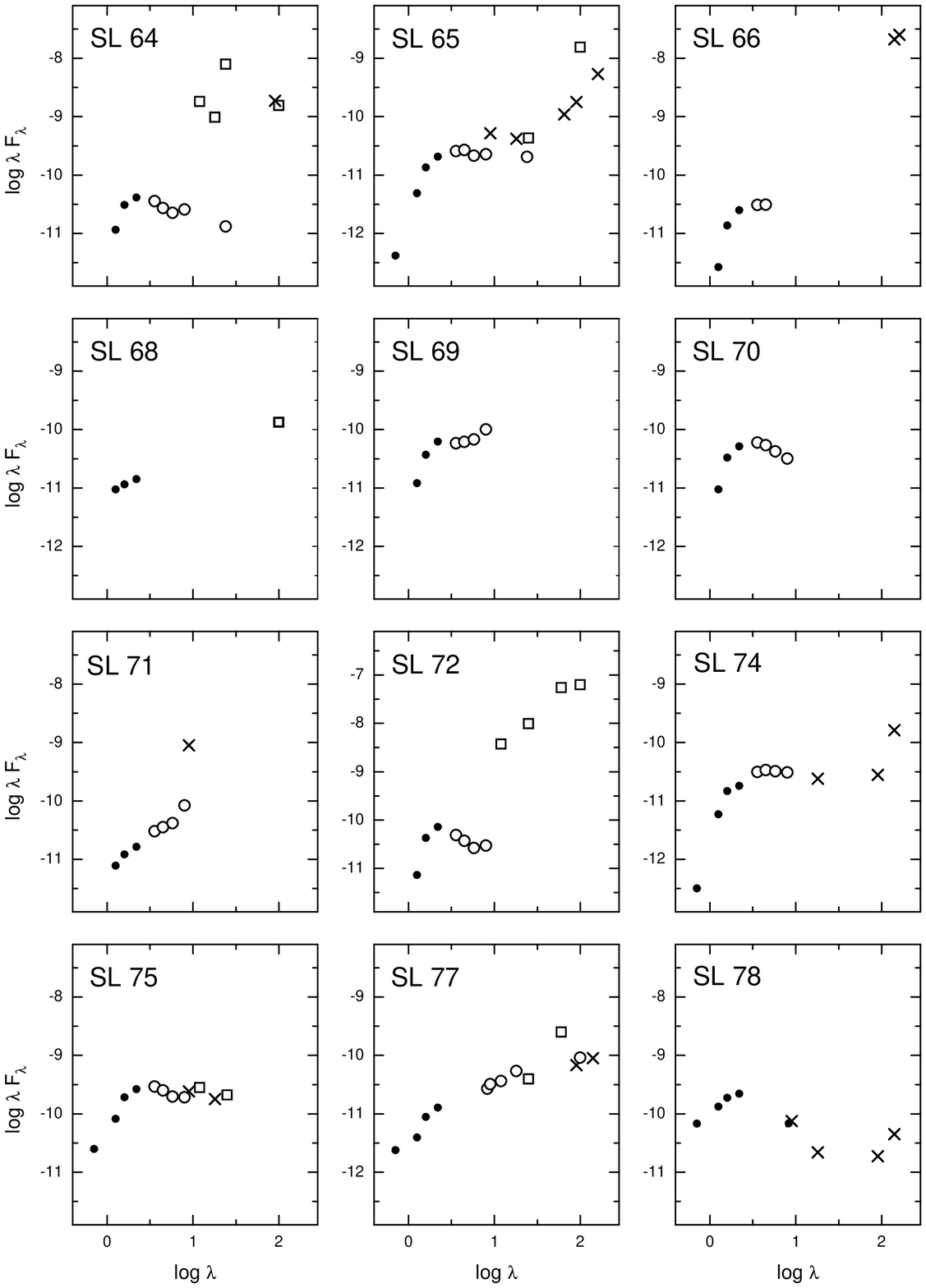,width=124mm,clip=}}
\captionc{1}{Continued}
}
\newpage

\vbox{
\centerline{\psfig{figure=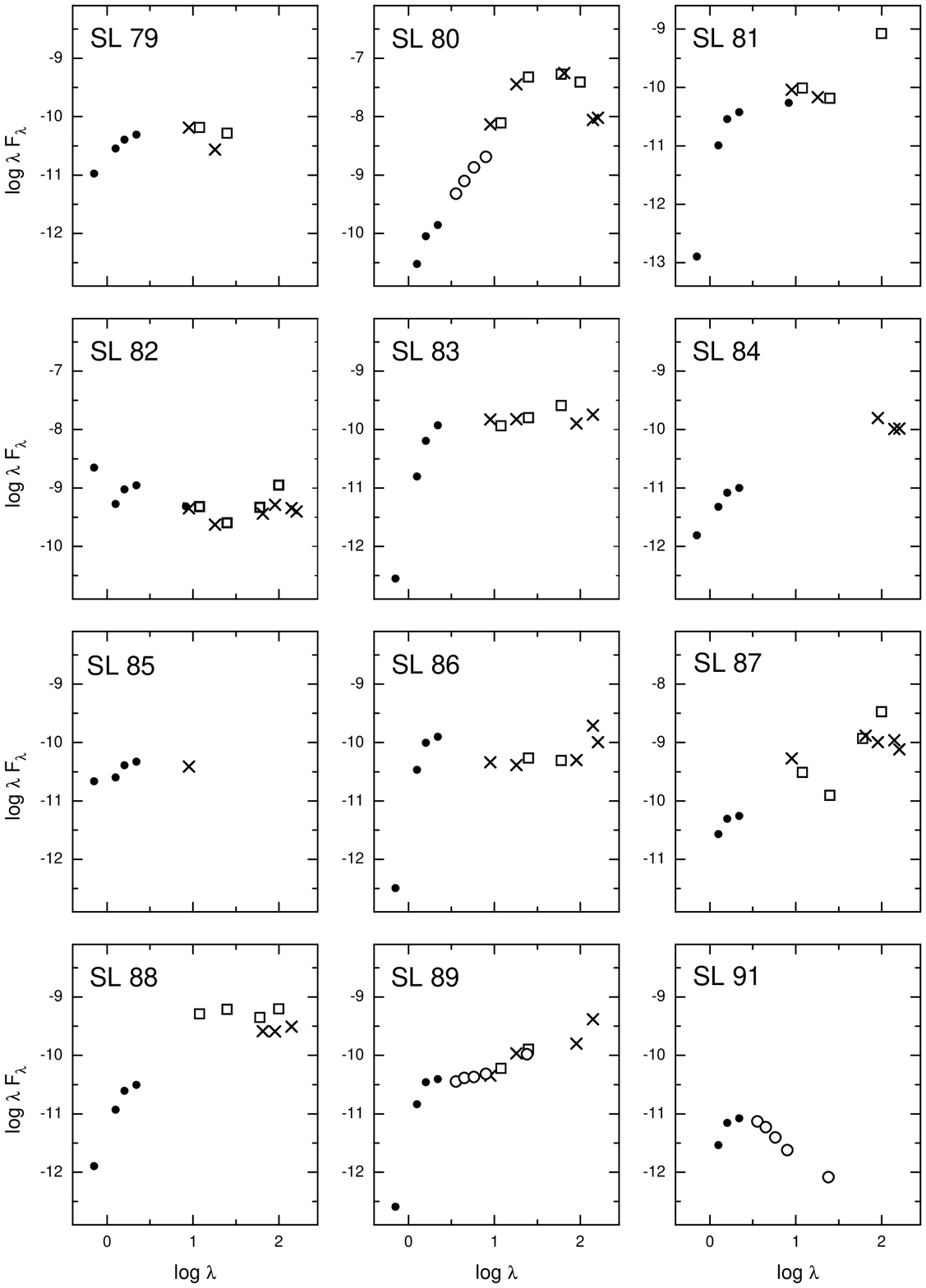,width=124mm,clip=}}
\captionc{1}{Continued}
}
\newpage

\vbox{
\centerline{\psfig{figure=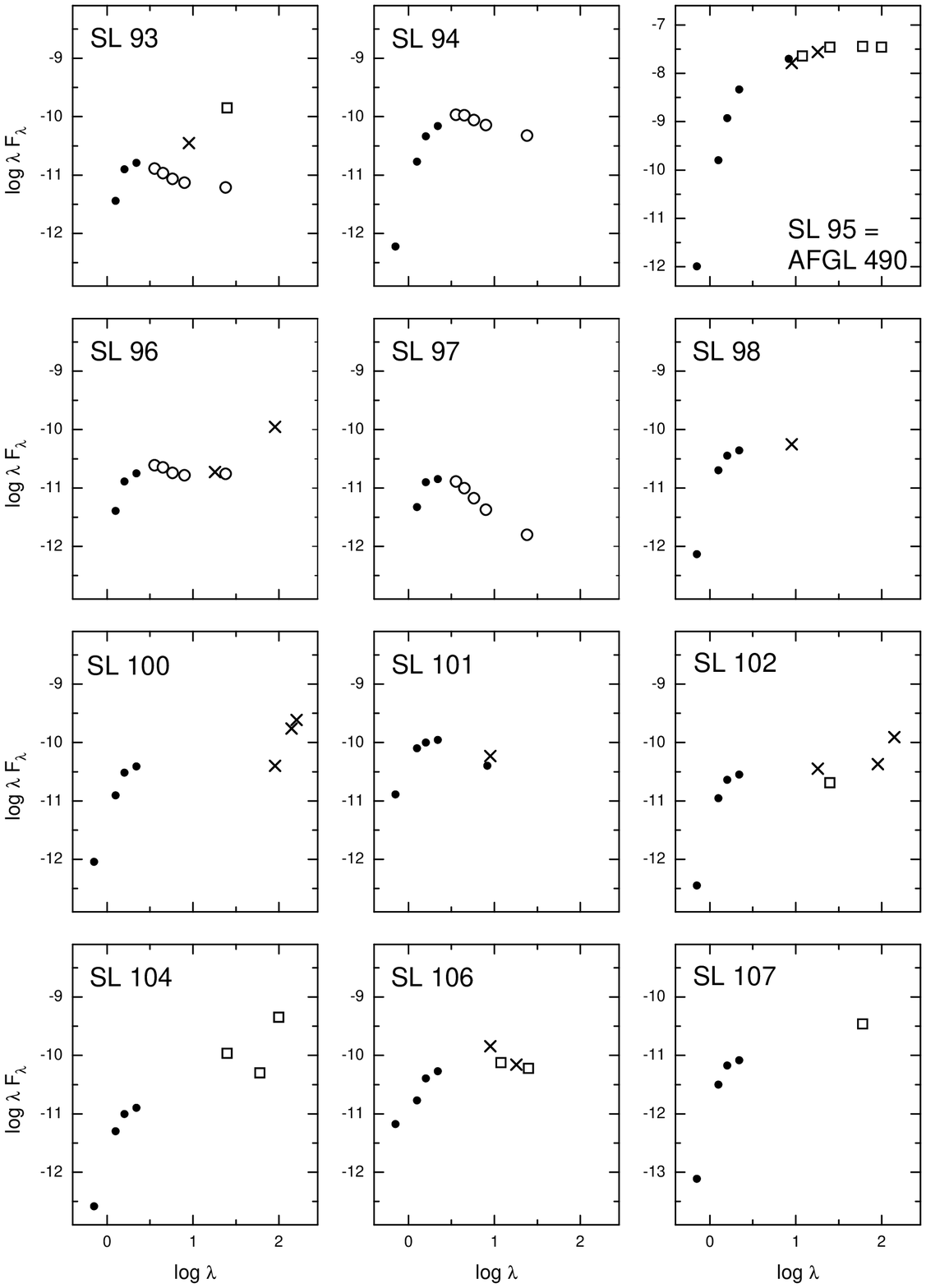,width=124mm,clip=}}
\captionc{1}{Continued}
}
\newpage

\vbox{
\centerline{\psfig{figure=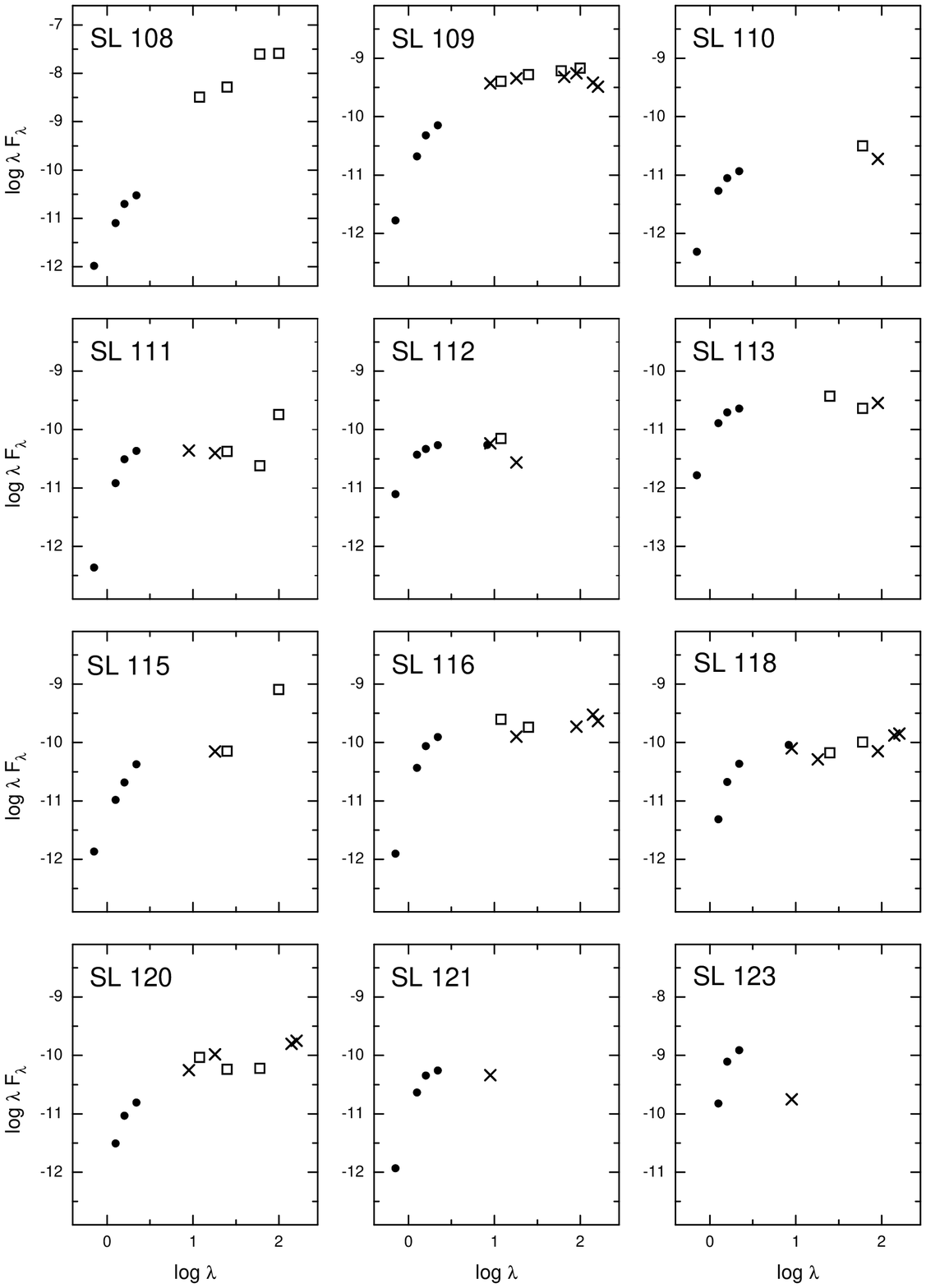,width=124mm,clip=}}
\captionc{1}{Continued}
}
\newpage

\vbox{
\centerline{\psfig{figure=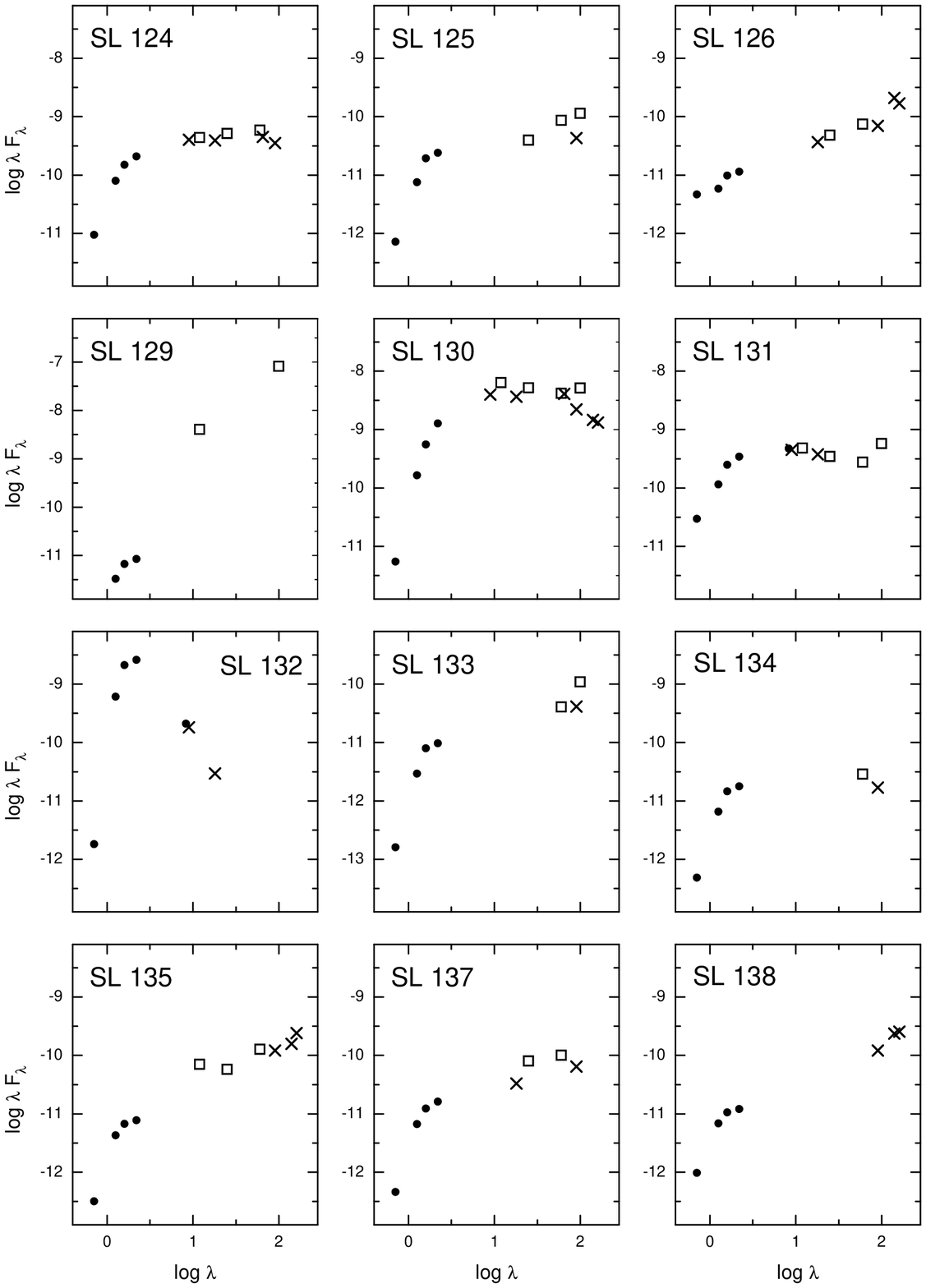,width=124mm,clip=}}
\captionc{1}{Continued}
}
\newpage

\vbox{
\centerline{\psfig{figure=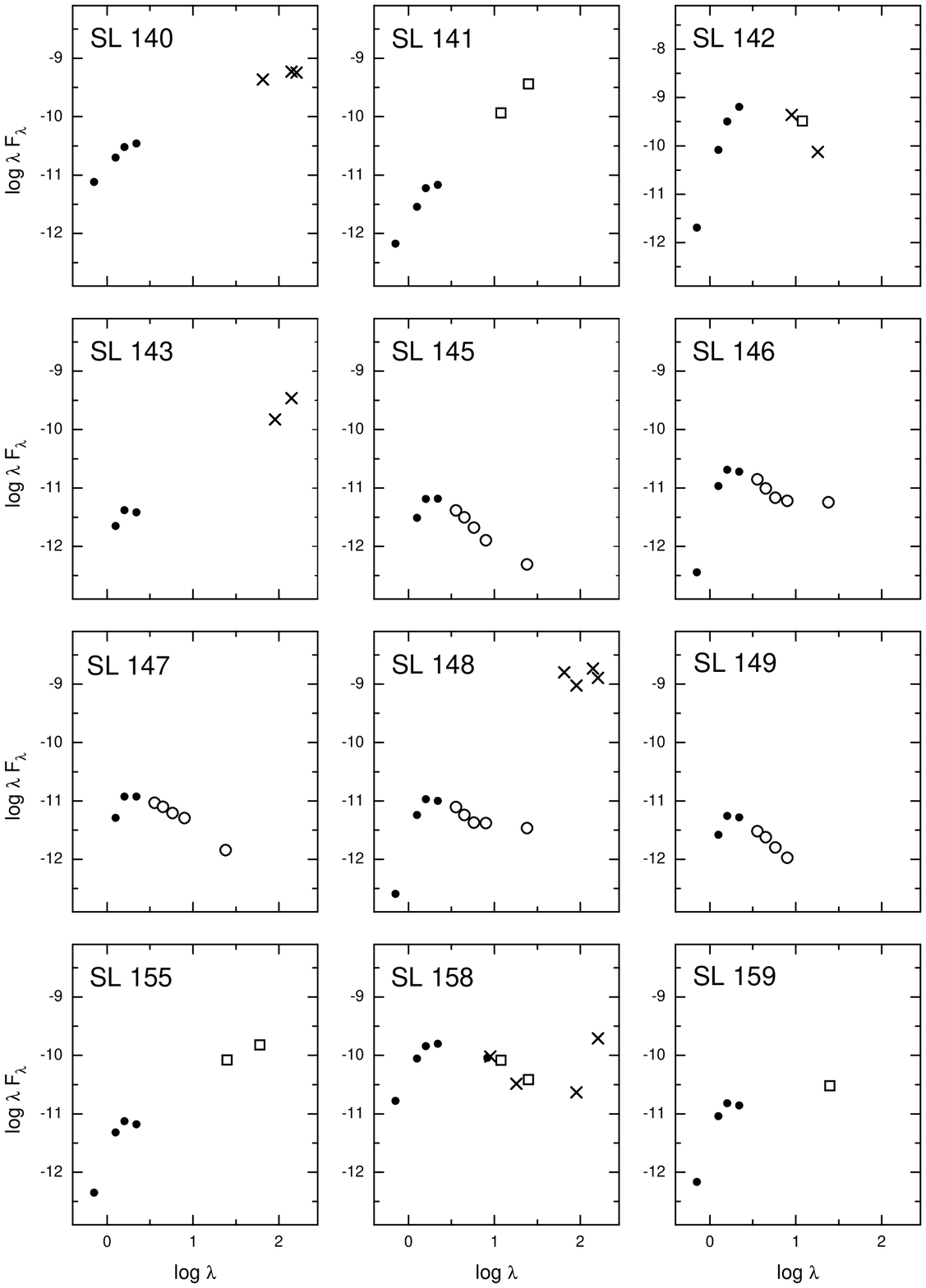,width=124mm,clip=}}
\captionc{1}{Continued}
}
\newpage

\vbox{
\centerline{\psfig{figure=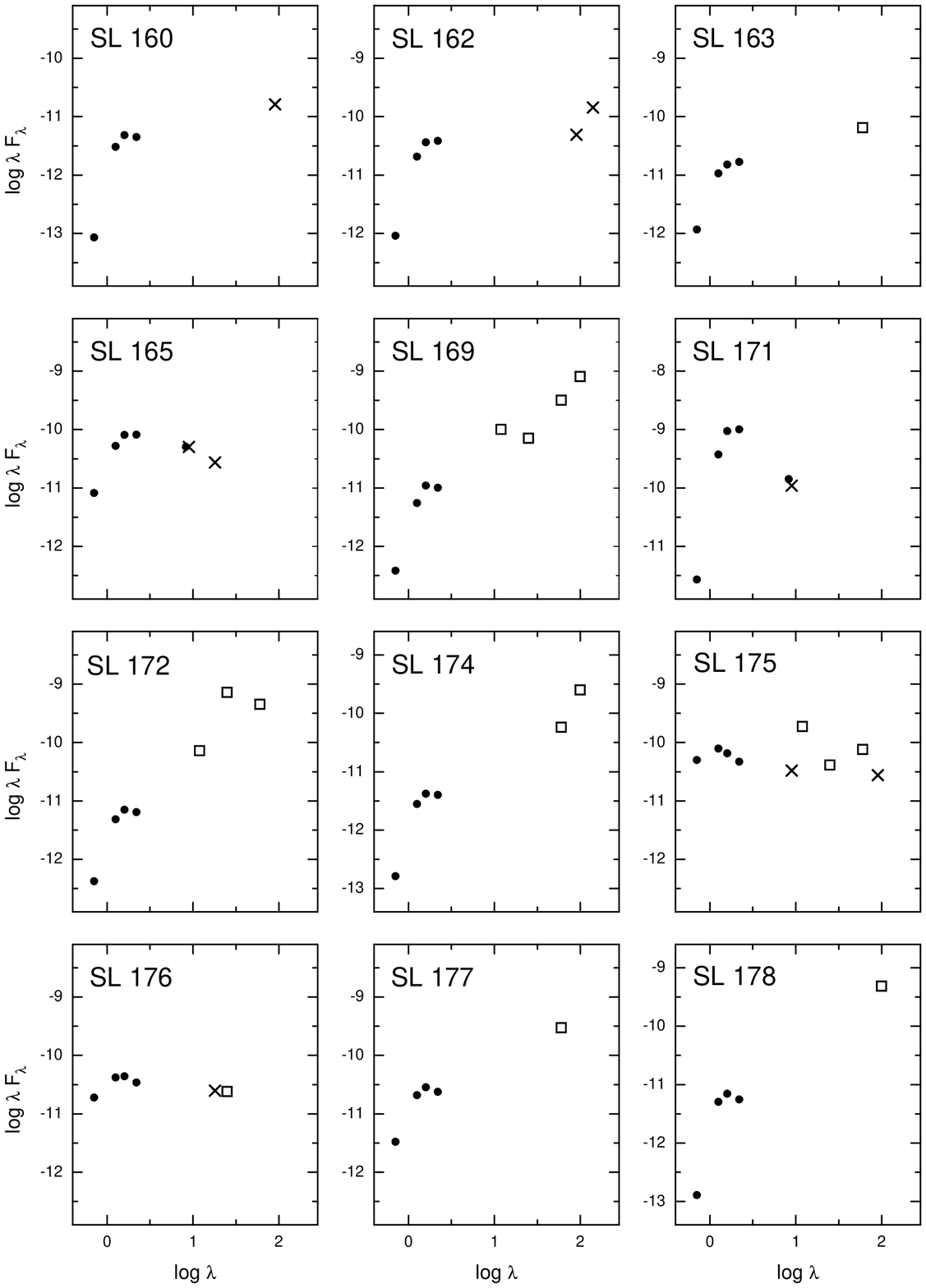,width=124mm,clip=}}
\captionc{1}{Continued}
}
\newpage

\vbox{
\centerline{\psfig{figure=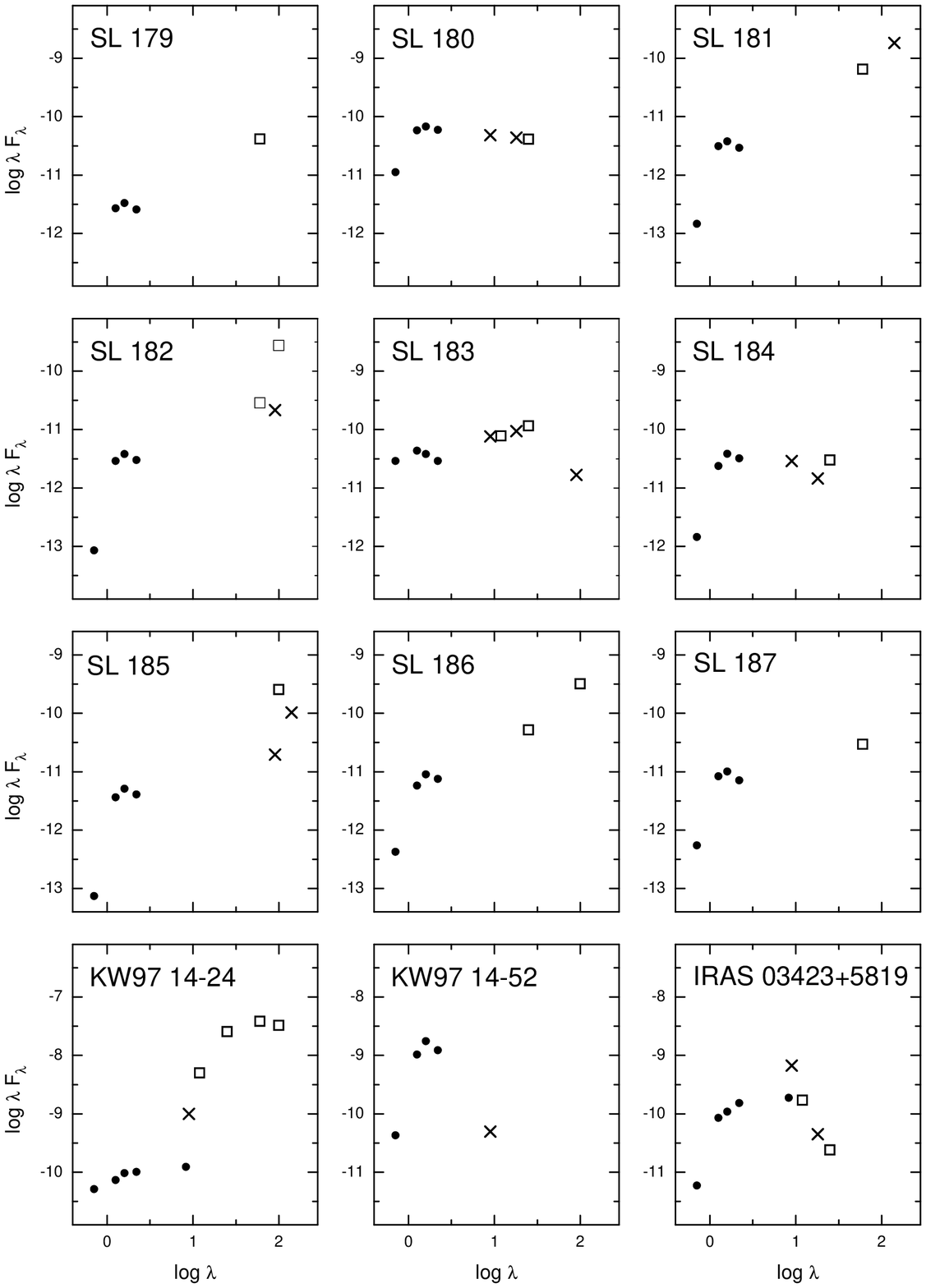,width=124mm,clip=}}
\captionc{1}{Continued}
}
\newpage

\vbox{
\centerline{\psfig{figure=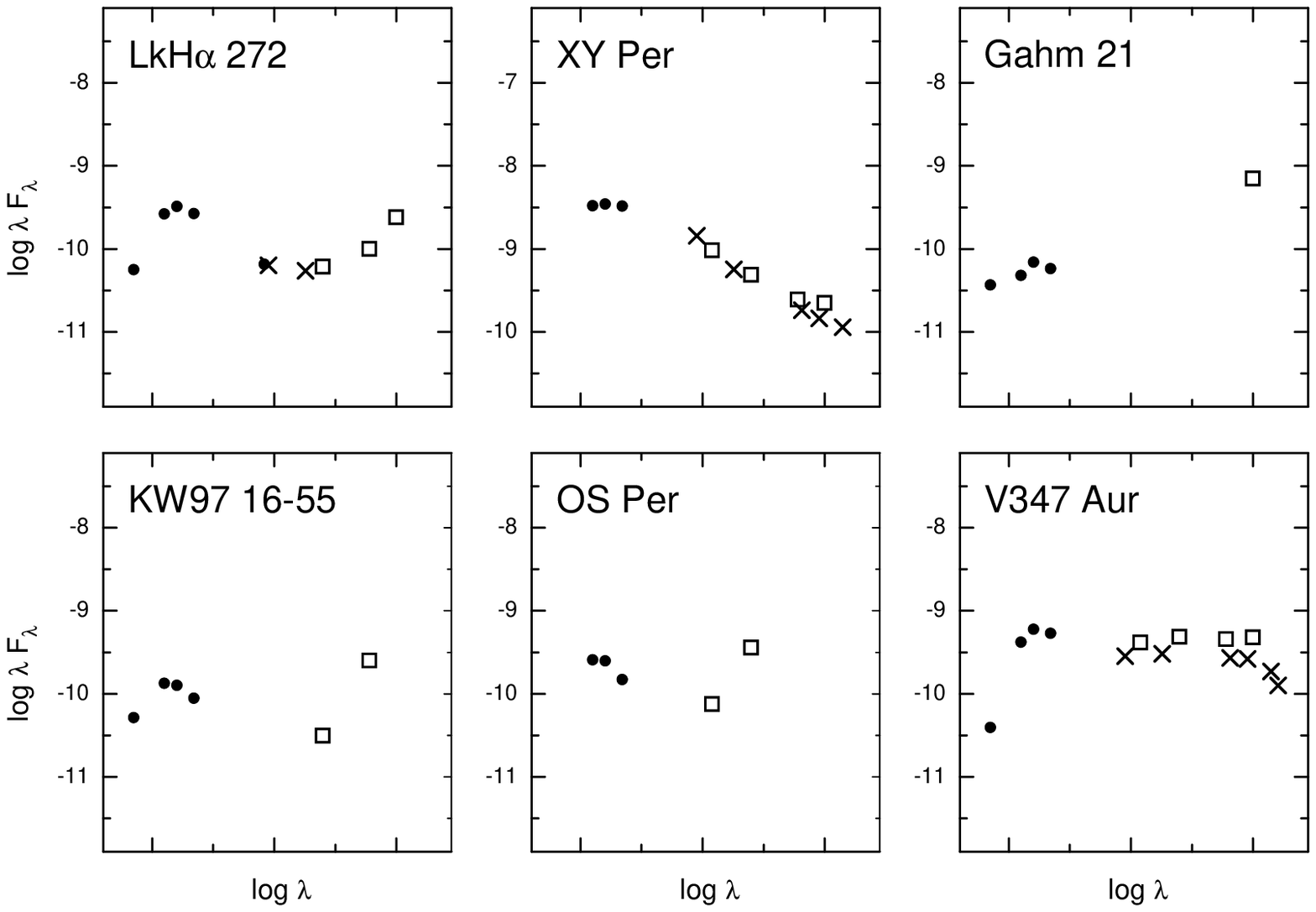,width=124mm,clip=}}
\vspace{-5mm}
\captionc{1}{Continued}
}
\vskip4mm

\noindent were inspected in the blue, red and infrared images of DSS2
given in the Internet's Virtual Telescope SkyView.  The results are
described in the following section.

According to Pollo et al.  (2010), stars and galaxies can be separated
using some flux vs. color and color vs. color diagrams constructed only
from the AKARI FIS fluxes.  However, their method cannot be applied in
our case since for most of the SL objects reliable fluxes are available
not in all FIS passbands.

A part of the extended objects discovered by infrared imaging may be
so-called ultracompact H\,II regions, the newly born massive stars
hidden in dense dust cocoons (Churchwell 2002).  SEDs of these
objects rise steeply with increasing $\lambda$ and have their intensity
maxima at 100 $\mu$m.

In Figure 2 we show the $J$--$H$ vs.\,$H$--$K_s$ diagram for SL objects
from Table 1:  dots denote YSOs without extended sources, circles denote
19 YSOs with a nearby extended infrared source and crosses denote 14
YSOs with a nearby galaxy.  It is evident that all the three types of
YSOs do not show significant differences in their location.  This means
that either the radiation from the extended sources is relatively faint
shortward of 2.5 $\mu$m or its effect is similar to that of the
interstellar and circumstellar reddening.  The concentration of points
at $H$--$K_s$ = 1.0 is the selection effect originating from a search
for YSOs in Paper II.

\begin{figure}[!t]
\vbox{
\centerline{\psfig{figure=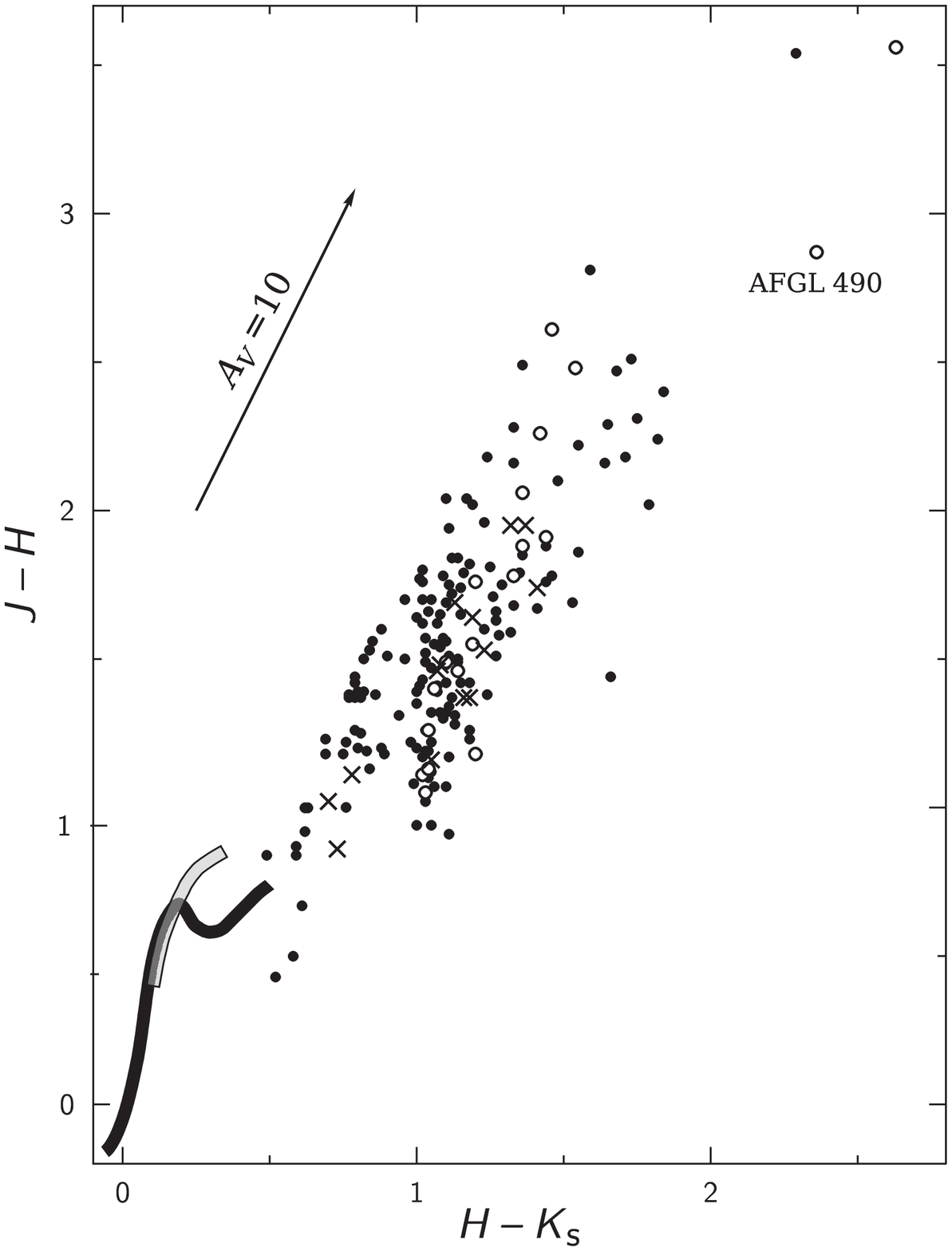,width=78mm,clip=}}
\captionb{2}{The $J$--$H$ vs.\,$H$--$K_s$
diagram for the verified YSOs. Dots denote objects without extended
sources, circles denote objects with nearby extended infrared sources
and crosses denote objects with nearby galaxies. The black and grey
curves denote the main sequence and G8--K--M giants, respectively.}
}
\end{figure}

\sectionb{4}{DISCUSSION}

In this section we describe SEDs of the SL objects belonging to
different star-forming regions or dust/molecular clouds.

\subsectionb{4.1}{Objects in W3A (IC 1795) and the dust cloud TGU\,879}

This group contains the following SL objects:  7, 9, 10, 11, 13, 15, 16,
18, 20, 21, 23, 25, 27, 30, 31, 34, 35, 36, 37, 38, 39.  Most of these
are located in the W3A (IC 1795) H\,II region, some are seen in the
direction of the dust cloud TGU\,879 located south of it.  Both the
H\,II region and the dust cloud belong to the Perseus arm (see Paper
II).  Some of these objects were discovered as YSOs and discussed by
Elmegreen (1980), Kerton et al.  (2001) and Kerton \& Brunt (2003).
Most of these objects exhibit SEDs typical for Class I YSOs.  The
objects SL 15, 20, 21 and 31 may belong to Class II and SL 18, 30, 34
and, possibly, 37 to Class III.  The data for the object AKARI-FIS-V1
J0224555+620941, located at 25.6\arcsec\ from SL\,18, contradict the
Spitzer data and may belong to a different object.  A similar situation
is with the object AKARI-FIS-V1 J0227184+620030 at 26.8\arcsec\ from
SL\,37.  The SEDs of some of the objects are extremely steep (SL 23, 25,
27, 35 and 36).  This should be an indication of heavy interstellar and
circumstellar reddening which makes the magnitudes $F$, $J$, $H$ and
$K_s$ very faint.  This conclusion is also supported by the absence of
$F$ magnitudes for these stars in the DSS2 survey (they are fainter than
20.5 mag).  Two of the objects, SL\,21 and SL\,37, were confirmed as
YSOs in Paper IV, their spectral types are G0e and A5e.  According to
2MASX, the object SL\,10 at a distance of 0.2\arcmin\ has a galaxy.  Two
extended 2MASX objects are located at SL\,23 (0.7\arcmin) and at SL\,37
(0.4\arcmin).

\subsectionb{4.2}{Objects in the dust cloud TGU\,878 near SU Cas}

Six SL objects (Nos. 3, 4, 5, 6, 8 and 14) are located in the dust cloud
TGU\,878 at $\ell$\,$\approx$\,132.5\degr, $b$\,$\approx$\,+9\degr,
belonging to the Cam OB1 association located at a distance of 900 pc.
The same cloud contains the cepheid SU Cas.  The SEDs of the suspected
YSOs are of Classes I and II.  The object SL\,8 = IRAS 02470+6901
according to NED may be a galaxy.

\subsectionb{4.3}{Objects in W4}

Five SL objects, Nos. 40, 41, 43, 44 and 48, are projected on the W4
H\,II region of the Perseus arm.  However, SL\,40, according to 2MASX,
can be a galaxy centered at 0.3\arcmin.  SL\,44 = IRAS 02245+6115 at
0.2\arcmin\ and SL\,48 = IRAS 02327+6019 at 0.4\arcmin\ have extended
2MASX sources which might be related to the infrared clusters identified
by Bica et al.  (2003a).

\subsectionb{4.4}{Objects near W5}

This group is located in the dust clouds TGU\,879 and TGU\,912 along the
northern edge of the W5 nebula and contains 22 SL objects from Figure 1:
50, 51, 52, 53, 54, 55, 56, 57, 59, 60, 61, 62, 63, 64, 65, 66, 69, 70,
71, 72, 74 and 75.  Some of these objects were identified by Carpenter
et al.  (2000), Kerton \& Brunt (2003), Karr \& Martin (2003a,b), Bica
et al.  (2003a,b).  The SEDs of these objects are consistent with YSOs
of Classes I and II.  The objects SL 55, 56, 57, 60, 63, 64, 66, 69, 70,
71 and 72 have extended counterparts at distances 0.2\arcmin\ to
1\arcmin\ listed in 2MASX.  Two of them, counterparts of SL\,57 and 60,
are galaxies.  For the object SL\,64, the Spitzer and IRAS/AKARI data
are in disagreement.  The star SL\,63 is a known pre-main-sequence
variable, LW Cas, of type INA and spectral class A0.  In Paper IV the
object SL\,75 was investigated spectroscopically and recognized as a T
Tauri type star of spectral class G0e with strong emissions in
H$\alpha$, O\,I and Ca\,II lines.  One more YSO, KW97\,14-24, which is
also located near the edge of W5, was investigated in Paper IV and found
to be of spectral class G5e with H$\alpha$ of {\it EW} = --20.6 \AA.
Nakano et al.  (2008) for this star find a similar value, {\it EW} =
--19.4 \AA.  The SED curve of this object is shown in Figure 1 (page
25).

\subsectionb{4.5}{Objects in LBN\,140.07+1.64}

Five SL objects, Nos. 79, 80, 83, 84 and 86, are located in the
direction of the faint nebula LBN\,140.07+1.64 in the Perseus arm, south
of Sh\,2-202 which is much closer to the Sun.  The object SL\,80 is
known as one of YSOs in the infrared cluster AFGL\,437.  Other objects
were identified by Kerton \& Brunt (2003), Karr \& Martin (2003a,b) and
Bica et al.  (2003a).  If these objects are pre-main-sequence stars,
their SEDs agree with Class II.  The spectrum of SL\,79 gives spectral
type G0e, with strong emissions in H$\alpha$, O\,I and Ca\,II (Paper
IV).  The emission in H$\alpha$ was also found in the IPHAS survey
(Witham et al. 2008).  SL\,80 = AFGL\,437S was classified as an object
of Class I based on the Spitzer/IRAC data (Kumar Dewangan \& Anandarao
2010).  The SED of SL\,80 in Figure 1 shows some disagreement between
the 2MASS + Spitzer and AKARI data.  Probably the AKARI data are
affected by other members of the AFGL\,437 cluster.

\subsectionb{4.6}{Objects in Sh\,2-202}

The faint nebula Sh\,2-202, belonging to the Cam OB1 SFR, overlaps
several smaller emission features in the Perseus arm (Karr \& Martin
2003a).  In this direction, two our objects, SL\,81 and SL\,82, are
seen.  The spectrum of SL\,82 gives the spectral type G5e, with strong
emission in H$\alpha$ (Paper IV).  The object is also known as IRAS
03134+5958 and CPM\,7 (Campbell et al. 1989).  The object is included in
the catalog of emission-line stars of the Orion population by Herbig \&
Bell (1988).  The SED of this object is rather peculiar.

\subsectionb{4.7}{Objects in the dust cloud TGU\,942 in the vicinity
of AFGL\,490}

In Papers II and III, 60 possible YSOs were identified in the area of
3\degr\,$\times$\,3\degr\ with the center at $\ell$ = 142.5\degr, $b$ =
+1.0\degr, covering the densest part of the dust cloud TGU\,942.  The
SEDs for 39 of them are shown in Figure 1. Their SL numbers are from 89
to 102, 107 and from 143 to 187.  We shall describe their properties,
distributing them into three groups according to their $H$--$K_s$
values.

The group of 12 objects with $H$--$K_s$\,$\geq$\,1.0, SL numbers between
89 and 107.  The objects SL 89, 95, 100 and 107 belong to Class I. The
objects SL 93, 94, 96, 98, 101 and 102 belong to Class II.  The objects
SL\,91 and 97 are of Class III.  The object SL\,95 = AFGL\,490 is one of
the well known YSOs of Class I, a protostar of spectral class B. Its
spectrum was described in Paper IV, together with the spectrum of
SL\,101 of spectral class G0e.  Both objects have strong emissions in
H$\alpha$, O\,I and Ca\,II lines.  One more YSO, IRAS 03243+5819 = 2MASS
J03281460+5829374, located 20\arcmin\ from AFGL\,490, was classified in
Paper IV as A-type star with H$\alpha$ line filled by emission.  It
shows a peculiar SED (see Figure 1, p.\,25).  The object SL\,103, not
present in Figure 1, was found to be a H$\alpha$ emission star in the
IPHAS survey.  According to 2MASX, the object SL\,94 has an extended
infrared source at a distance of 0.4\arcmin.

The group of 17 objects with $H$--$K_s$ between 0.75 and 1.00, SL
numbers between 143 and 174.  The objects SL 143, 155, 160, 163, 169,
172 and 174 belong to Class I, SL 146, 148, 158, 159, 162, 165 to Class
II, and SL 145, 147, 149, 171 to Class III.  For SL\,148, the data from
Spitzer and AKARI are in obvious contradiction.  Probably the
2MASS+Spitzer object and the AKARI object, separated by 28\arcsec, are
different objects.  The objects SL\,158 (A0e), SL\,163 (F:e) and SL\,165
(G5e) were investigated spectroscopically (Papers IV and V).  The
spectra of two more stars of this group, SL\,144 (G0e) and SL\,153
(F0e), were obtained (Paper V), but no infrared data for these objects
are available.  In SL\,161, H$\alpha$ emission was found by the IPHAS
survey.  For the objects SL\,169 and SL\,174, the identification with
the IRAS sources is doubtful ($>$\,1\arcmin).  According to 2MASX, a
small galaxy is located at 0.4\arcmin\ from SL\,172.

The group of 13 objects with $H$--$K_s$ between 0.50 and 0.75, SL
numbers between 175 and 187.  The objects SL 177, 178, 179, 181, 182,
185, 186 and 187 belong to Class I. The objects SL 175, 176, 180, 183
and 184 belong to Class II.  In Paper V the spectra for the following
objects were investigated:  SL\,175 (F0e), SL\,176 (G0e), SL\,177 (A2e),
SL\,183 (A2e), SL\,184 (F:e).  In SL\,177, H$\alpha$ emission was
found by the IPHAS survey.  For the objects SL\,175 and SL\,185, the
IRAS identification is doubtful ($>$\,1\arcmin).

\subsectionb{4.8}{Objects in the remaining part of the dust cloud
TGU\,942}

Figure 1 contains eight more objects in the TGU\,942 cloud scattered
over nearly 3\degr\ along the Galactic longitude.  Most of these objects
are projected on the molecular clouds of the Perseus arm (see Paper II).
Four SL objects (104, 105, 106 and 108) are located in the faint
emission nebula Sh\,2-203 and belong to the infrared cluster BDSB\,59
(Bica et al. 2003b).  The SEDs of three of these objects are shown in
Figure 1. According to the 2MASX catalog, SL\,106 is only by 0.4\arcmin\
from a galaxy.  The object SL\,109 is the well-known Class I YSO CPM\,8
(Campbell et al. 1989).  The remaining four objects, SL 110, 111, 112
and 113 are likely to be YSOs of Class II.

\subsectionb{4.9}{Objects in the vicinity of the Sh\,2-205 emission
nebula}

Four SL objects are located close to the H\,II region Sh\,2-205.  The
SEDs of two of them, SL\,115 and SL\,116, are shown in Figure 1:  both
seem to be of Class II.  Probably the stars belong to the infrared
cluster FSR\,655 (Froebrich et al. 2007).  H$\alpha$ emission in SL\,116
was found in the IPHAS survey.  In the same area, eight emission-line
stars discovered by G. Gahm and investigated spectroscopically in Papers
IV and V are located.  The SED of one of them, Gahm 21 (spectral type
Ke), is shown in Figure 1 (p.\,26), admitting that IRAS\,03561+5123 with
the position 33\arcsec\ away is the same object.  Other Gahm objects are
absent in the IRAS, MSX and AKARI databases.

\subsectionb{4.10}{Objects within the dust circle centered on the
open cluster NGC\,1528}

In Paper I, a circle of dust clouds with a diameter of $\sim$\,8\degr,
centered at ($\ell$,\,$b$) = (152\degr, +0.5\degr), close to the open
cluster NGC\,1528, was described.  Within this circle, Paper II lists 20
SL objects (from SL\,118 to SL\,137) but only 15 of these have SEDs
shown in Figure 1. The objects SL 126, 129, 133, 135 and 137 belong to
Class I, objects SL 118, 120, 121, 124, 125, 130, 131 and 134 to Class
II and objects SL 123 and 132 to class III.  The best known YSO is
SL\,130 = CPM\,12 in the nebula Sh~2-209 (Campbell et al. 1989).
Objects SL 123, 129 and 130 may be related to the infrared clusters
BDSB\,61 and BDSB\,65 (Bica et al. 2003b).  SL\,124 is known as a T
Tauri type star (GLMP\,49, Garc\'ia-Lario et al. 1997).  Objects SL 126
and 137 may be overlapped by images of the nearby galaxies.

Four SL objects, 138, 140, 141 and 142, are located at the extension of
dust clouds of the afore-mentioned circle, close to the dust clouds
TGU\,1045 and TGU\,1056.  The first three objects exhibit SEDs of Class
I and SL\,142 of Class III.  CO radial velocities show that the object
SL\,140 belongs to the Perseus arm (Wouterloot \& Brand 1989).

\subsectionb{4.11}{Other objects}

SL\,1 = IRAS 02081+6225. According to the NED and 2MASX databases this
object is a galaxy.

SL\,49 = IRAS 02475+6156.  By infrared imaging Iwata et al. (1997)
find this object to be a possible galaxy.

SL\,58. Extended source in the DSS2 blue, red and infrared images;
a possible galaxy.

SL\,68 = IRAS 03074+6211. Extended or binary source in DSS2 blue, red
and infrared images.

SL\,77 and 78.  These two objects, separated by 3.2\arcmin, are located
in the direction of the dark cloud TGU\,931 which belongs to dust layer
of the Cam OB1 association.  SL\,77 = 2MASS J02512410+5542038 can be
related to IRAS 02476+5529 at 72\arcsec\ or AKARI-FIS-V1 J0251208+554217
at 31\arcsec.  In the IPHAS survey it is recognized as a H$\alpha$
emission star.  The object SL\,78 = 2MASS J02514696+5542014 was
classified as F5e in Paper IV.  If it is related to the object
AKARI-FIS-V1 J0251409+554223 at a distance of 56\arcsec, the YSO should
belong to Class II.

SL\,85 = 2MASS J03300237+6125473.  The object is seen in the direction
free of dark clouds.  Another object (2MASS J03300208+6125565) of oblong
shape and with similar brightness in the red and near infrared is
located at 9\arcsec\ north.  This second object is fainter than SL\,85
both in the blue and in the {\it J, H, K}$_s$ magnitudes.  The
AKARI-IRC-V1 J0330022+612547 flux at 9 $\mu$m is available.  The objects
may be extragalactic.

SL\,87 and SL\,88.  These two objects of Class I are located in the
southern branch of the dark cloud TGU\,942 towards the small emission
nebula LBN 140.77-1.42 (Karr \& Martin 2003a) in the Perseus arm.
H$\alpha$ emission in the spectrum of SL\,88 was found by the IPHAS
survey.

\sectionb{5}{CONCLUSIONS}

In the previous papers in this series (Papers II and III), 187 objects
(designated as SL objects) located in the Milky Way between Galactic
longitudes 132--158\degr\ and latitudes $\pm$\,12\degr, were suspected
as pre-main-sequence stars or YSOs on the grounds of their 2MASS, IRAS
and MSX infrared colors.  The Spitzer and AKARI data release gave us the
possibility to construct infrared spectral energy distributions for more
objects and to verify their evolutionary status.  In this study we
constructed the SEDs between 0.7~$\mu$m and 160~$\mu$m for 141 objects
of our sample, taking the data from the GSC\,2.3.2, 2MASS, IRAS, MSX,
Spitzer and AKARI catalogs.  The analysis of the SEDs of these objects,
combined with their surface distribution, leads to the conclusion that
about 45\%, 41\% and 14\% of them can be YSOs of Classes I, II and III,
respectively.  Most of these objects are heavily reddened by
interstellar and circumstellar dust what makes them quite faint in the
red and near infrared spectral regions.  About 30 of SL objects have
extended 2MASX sources within 1\arcmin, which can affect photometric
measurements of the point-like sources.  Such extended sources usually
are small ionized regions, clusterings of faint infrared stars or
galaxies.

Star-forming dusty galaxies contaminate SEDs of YSOs, rising the
intensity at 100~$\mu$m due to dust emission.  This makes the SED of an
YSO similar to that of the Class I object.  The presence of a galaxy can
be identified either by inspecting optical or near infrared images in
the vicinity of YSO or by spectral observations.  We identified 13
galaxies in the vicinity of SL objects by inspecting blue, red and
infrared images of the DSS2 survey and checking the NED database as well
as 2MASX and SDSS catalogs.  We cannot exclude that more SEDs remain
contaminated by nearby unidentified galaxies.

The concentration of YSOs in close groups, clusters and gas/dust clouds
provides an additional statistical criterion for their separation from
galaxies.  However, the concentration of infrared objects towards dense
clouds alone does not mean that we have a group of objects located close
to each other in space.  As it was shown in Paper III, distant heavily
reddened K and M giants also demonstrate a tendency to concentrate in
the direction of the densest dust clouds, thus imitating clustering of
real YSOs.  Therefore, for the identification of embedded groups and
clusters of YSOs, photometry in the mid- and far-infrared and spectral
observations become essential.


\newpage

\thanks {This research includes observations with AKARI, a JAXA
project with the participation of ESA.  The use of the 2MASS, IRAS, MSX,
Spitzer, SkyView and Simbad databases is acknowledged.  We are
thankful to Stanislava Barta\v{s}i\={u}t\.e, Vygandas Laugalys
and Edmundas Mei\v{s}tas for their help preparing the paper.}

\References

\refb Abazajian K. N., Adelman-McCarthy J. K., Agueros M. A. et al.
2009, ApJS, 182, 543

\refb Bica E., Dutra C. M., Barbuy B. 2003a, A\&A, 397, 177

\refb Bica E., Dutra C. M., Soares J., Barbuy B. 2003b, A\&A, 404, 223

\refb Campbell B., Persson S. E., Matthews K. 1989, AJ, 98, 643

\refb Carpenter J. M., Heyer M. H., Snell R. L. 2000, ApJS, 130, 381

\refb Churchwell E. 2002, ARA\&A, 40, 27

\refb Corbally C. J., Strai\v{z}ys V. 2008, Baltic Astronomy, 17, 21
(Paper IV)

\refb Corbally C. J., Strai\v{z}ys V. 2009, Baltic Astronomy, 18, 1
(Paper V)

\refb Dale D. A., Helou G. 2002, ApJ, 576, 159

\refb Devriendt J.\,E.\,G., Guiderdoni B., Sadat R. 1999, A\&A, 350,
381

\refb Dobashi K., Uehara H., Kandori R., Sakurai T., Kaiden M.,
Umemoto T., Sato F. 2005, PASJ, 57, S1

\refb Elmegreen D. M. 1980, ApJ, 240, 846

\refb Froebrich D., Scholz A., Raftery C. L. 2007, MNRAS, 374, 399

\refb Garc\'ia-Lario P., Manchado A., Pych W., Pottasch S. R. 1997,
A\&AS, 126, 479

\refb Gutermuth R. A., Megeath S. T., Myers P. C. et al. 2009, ApJS,
184, 18

\refb Herbig G. H., Bell K. R. 1988, Lick Obs. Bull., No. 1111

\refb Ishihara D., Onaka T., Kataza H. et al. 2010,
{\it The AKARI/IRC Mid-Infrared All-Sky Survey}, A\&A, 514, A1

\refb Iwata I., Nakanishi K., Takeuchi T. et al. 1997, PASJ, 49, 47

\refb Karr J. L., Martin P. G. 2003a, ApJ, 595, 880

\refb Karr J. L., Martin P. G. 2003b, ApJ, 595, 900

\refb Kataza H., Alfageme C., Cassatella A. et al. 2010, {\it AKARI-IRC
Point Source Catalogue Release Note, Version 1.0}\\
http://www.ir.isas.jaxa.jp/AKARI/Observation/PSC/Public/

\refb Kerton C. R., Brunt C. M. 2003, A\&A, 399, 1083

\refb Kerton C. R., Martin P. G., Johnstone D., Ballantyne D. R. 2001,
ApJ, 552, 601

\refb Koenig X. P., Allen L. E., Gutermuth R. A. et al. 2008, ApJ, 688,
1142

\refb Kumar Dewangan L., Anandarao B. G. 2010, MNRAS, 402, 2583

\refb Lada C. J. 1987, in {\it Star Forming Regions} (IAU Symp. 115),
eds. M. Peimbert \& J. Jugaku, Reidel Publ. Comp., Dordrecht, p.\,1

\refb Lasker B. M., Lattanzi M. G., McLean B. J. et al. 2008, AJ, 136,
735; CDS Catalog I/305, version GSC\,2.3.2

\refb Luhmann K. L., Whitney B. A., Meade M. R. et al. 2006, ApJ, 647,
1180

\refb Murakami H., Baba H., Barthel P. et al. 2007, PASJ, 59, S369

\refb Nakano M., Sugitani K., Niwa T. et al. 2008, PASJ, 60, 739

\refb NED, 2010, http://nedwww.ipac.caltech.edu/

\refb Pollo A., Rybka P., Takeuchi T. T. 2010, A\&A, 514, A3

\refb Reach W. T., Megeath S. T., Cohen M. et al. 2005, PASP, 117, 978

\refb Rebull L. M., Padgett D. L., McCabe C.-E. et al. 2010, ApJS, 186,
259

\refb Rieke G. H., Blaylock M., Decin L. et al. 2008, AJ, 135, 2245

\refb Robitaille T. P., Whitney B. A., Indebetouw R. et al. 2006, ApJS,
167, 256

\refb Robitaille T. P., Whitney B. A., Indebetouw R., Wood K. 2007,
ApJS, 169, 328

\refb Ruch G. T., Jones T. J., Woodward C. E. et al. 2007, ApJ, 654, 338

\refb Sajina A., Scott D., Dennefeld M. et al. 2006, MNRAS, 369, 939

\refb Shirahata M., Matsuura S., Hasegawa S. et al. 2009, PASJ, 61, 737

\refb Skrutskie M. F., Cutri R. M., Stiening R. et al. 2006, AJ, 131,
1163 (2MASX catalog)

\refb Strai\v{z}ys V. 1992, {\it Multicolor Stellar Photometry},
Pachart Publ.  House, Tucson, Arizona (available in pdf format at
http://www.itpa.lt or http://www.tfai.vu.lt)

\refb Strai\v{z}ys V., Laugalys V. 2007a, Baltic Astronomy, 16, 167
(Paper I)

\refb Strai\v{z}ys V., Laugalys V. 2007b, Baltic Astronomy, 16, 327
(Paper II)

\refb Strai\v{z}ys V., Laugalys V. 2008a, Baltic Astronomy, 17, 1
(Paper III)

\refb Strai\v zys V., Laugalys V. 2008b, {\it Young Stars and Clouds in
Camelopardalis}, in {\it Handbook of Star Forming Regions, vol.\,1. The
Northern Sky}, ed. B. Reipurth, ASP, p.\,294

\refb Whitney B. A., Wood K., Bjorkman J. E., Wolff M. J.. 2003a, ApJ,
591, 1049

\refb Whitney B. A., Wood K., Bjorkman J. E., Cohen M. 2003b, ApJ, 598,
1079

\refb Whitney B. A., Indebetouw R., Bjorkman J. E., Wood K. 2004, ApJ,
617, 1177

\refb Witham A. R., Knigge C., Drew J. E. et al. 2008, MNRAS, 384, 1277

\refb Wouterloot J.\,G.\,A., Brand J. 1989, A\&AS, 80, 149

\refb Yamamura I., Makiuti S., Ikeda N. et al. 2010,
{\it AKARI-FIS Bright Source Catalogue Release Note},\\
http://www.ir.isas.jaxa.jp/AKARI/Observation/PSC/Public/

\refb Zdanavi\v{c}ius J., Zdanavi\v{c}ius K., Strai\v{z}ys V. 2005,
Baltic Astronomy, 14, 31

\end{document}